\documentclass[a4paper,fleqn,usenatbib]{mnras}

\usepackage[T1]{fontenc}
\usepackage{ae,aecompl}

\usepackage{graphicx}	
\usepackage{amsmath}	
\usepackage{amssymb}	

\usepackage{color}
\usepackage[3D]{movie15}

\title[Stellar mass growth along the cosmic web]
	{Galaxy And Mass Assembly (GAMA): Stellar mass growth of spiral galaxies in the cosmic web}
\author[M. Alpaslan et al.]
	     {Mehmet Alpaslan$^1$\thanks{E-mail: mehmet.alpaslan@nasa.gov}, Meiert W. Grootes$^2$, Pamela M. Marcum$^1$, Cristina Popescu$^{3,4,2}$, \newauthor
	     Richard Tuffs$^2$, Joss Bland-Hawthorn$^5$, Sarah Brough$^6$, Michael J. I. Brown$^7$, \newauthor
	     Luke J. M. Davies$^8$, Simon P. Driver$^8$, Benne W. Holwerda$^9$, Lee S. Kelvin$^{10}$, \newauthor
	     Maritza A. Lara-L\'{o}pez$^{11}$, \'Angel R. L\'opez-S\'anchez$^{6,12}$, Jon Loveday$^{13}$, Amanda Moffett$^8$, \newauthor
	     Edward N. Taylor$^{14}$, Matt Owers$^{12,6}$, Aaron S. G. Robotham$^8$\\
$^1$NASA Ames Research Center, N232, Moffett Field, Mountain View, CA 94035, United States\\
$^2$Max Planck Institute fuer Kernphysik, Saupfercheckweg 1, 69117 Heidelberg, Germany\\
$^3$Jeremiah Horrocks Institute, University of Central Lancashire, PR1 2HE, Preston, UK\\
$^4$The Astronomical Institute of the Romanian Academy, Str. Cutitul de Argint 5, Bucharest, Romania\\
$^5$Sydney Institute for Astronomy, School of Physics A28, University of Sydney, NSW 2006, Australia\\
$^6$Australian Astronomical Observatory, PO Box 915, North Ryde, NSW 1670, Australia\\
$^7$School of Physics and Astronomy, Monash University, Clayton, Victoria 3800, Australia\\
$^8$International Centre for Radio Astronomy Research, 7 Fairway, The University of Western Australia, Crawley, Perth, Western Australia 6009, Australia\\
$^9$University of Leiden, Sterrenwacht Leiden, Niels Bohrweg 2, NL-2333 CA Leiden, The Netherlands\\
$^{10}$Astrophysics Research Institute, Liverpool John Moores University, IC2, Liverpool Science Park, 146 Brownlow Hill, Liverpool L3 5RF, UK\\
$^{11}$Instituto de Astronom\'{i}a, Universidad Nacional Aut\'{o}noma de M\'{e}xico, A.P. 70-264, 04510 M\'{e}xico, D.F., M\'{e}xico\\
$^{12}$Department of Physics and Astronomy, Macquarie University, NSW 2109, Australia\\
$^{13}$Astronomy Centre, University of Sussex, Falmer, Brighton BN1 9QH, UK\\
$^{14}$School of Physics, The University of Melbourne, Parkville, VIC 3010, Australia\\
}

\pubyear{2015}

\begin{document}
\label{firstpage}
\pagerange{\pageref{firstpage}--\pageref{lastpage}}
\maketitle

\begin{abstract}
We look for correlated changes in stellar mass and star formation rate along filaments in the cosmic web by examining the stellar masses and UV-derived star formation rates (SFR) of 1,799 ungrouped and unpaired spiral galaxies that reside in filaments. We devise multiple distance metrics to characterise the complex geometry of filaments, and find that galaxies closer to the cylindrical centre of a filament have higher stellar masses than their counterparts near the periphery of filaments, on the edges of voids. In addition, these peripheral spiral galaxies have higher specific star formation rates (SSFR) at a given mass. Complementing our sample of filament spiral galaxies with spiral galaxies in tendrils and voids, we find that the average SFR of these objects in different large scale environments are similar to each other with the primary discriminant in SFR being stellar mass, in line with previous works. However, the distributions of SFRs are found to vary with large-scale environment. Our results thus suggest a model in which in addition to stellar mass as the primary discriminant, the large-scale environment is imprinted in the SFR as a second order effect. Furthermore, our detailed results for filament galaxies suggest a model in which gas accretion from voids onto filaments is primarily in an orthogonal direction. Overall, we find our results to be in line with theoretical expectations of the thermodynamic properties of the intergalactic medium in different large-scale environments.
\end{abstract}




\section{Introduction}

When viewed at large scales, the distribution of galaxies in the Universe forms a vast network of interconnected filamentary structures, sheets, and clusters; all of which surround mostly empty voids. Commonly referred to as the cosmic web or the large-scale structure of the Universe, this arrangement of galaxies is a direct consequence of perturbations in the initial density field of matter shortly after the Big Bang evolving under the influence of gravity over cosmic time \citep{Zel'dovich1970,Shandarin1989,Bond1996}. A number of sophisticated algorithms now exist to identify and characterise large-scale structure in observed and simulated data sets \citep[e.g.][]{El-Ad1997,Doroshkevich2004,Aragon-Calvo2007,Colberg2007,Hahn2007,Platen2007,Neyrinck2008,Forero-Romero2009,Aragon-Calvo2010,Sousbie2011,Cautun2012,Alpaslan2013a,Tempel2014,Eardley2015}.

Many mechanisms that drive the evolution of a galaxy are sensitive to the environment in which it resides; particularly the local dark matter distribution \citep[e.g.][]{Brown2008,Yang2009,Zheng2009,Zehavi2011}. One must therefore take environment into account when studying the properties of galaxies; both locally (i.e. the properties of the group or pair in which the galaxy may reside) and globally (whether or not the galaxy resides in a filament or void). The relationship between a galaxy and its local environment is well studied, particularly in the context of its stellar population. Galaxies in groups and pairs often have star formation rates that deviate from those found in similar galaxies in the field \citep[e.g][]{Robotham2013a,Davies2015}. The degree to which large-scale structure impacts galaxy evolution is, however, less thoroughly studied. Recent work has shown that galaxies in voids have somewhat higher specific star formation rates compared to their counterparts in more dense environments \citep[e.g.][]{Rojas2004,Rojas2005,Kreckel2011,Kreckel2012,Ricciardelli2014}; though \citet{Penny2015} find that void galaxies with stellar masses $> 5 \times 10^9 M_{\odot}$ have largely ceased to form stars. \citet{Penny2015} also find that the colour-mass relationship of galaxies does not differ greatly between galaxies in voids and other environments. Concurrently, \citet{Fadda2008,Darvish2014} find that there is an enhancement in the fraction of star forming galaxies in filaments near large clusters.

Beyond the local Universe, there are strong indications that filaments served as conduits for galaxies to accumulate chemically evolved gas at $z \sim 3$; gas which is subsequently converted to young stars \citep{Kere2005,Gray2013} through enhanced star formation rates. A recent study by \citet{Snedden2016} indeed shows that in a suite of simulated filaments, star formation rates of galaxies at $z \sim 3$ close to the centre of a filament are lower compared to those in its periphery. Recently, \citet{Cautun2014} have shown that galaxies that reside in filaments at $z = 2$ remain in filaments or migrate into clusters by $z = 0$; therefore, galaxies that have formed in filaments experience a significant alteration of their masses and dynamics through filamentary flows, and then remain within such environments as they evolve. Such an evolutionary history is significantly different from a galaxy residing in an underdense void. In the local Universe, the effect of large-scale structure on galaxy stellar populations has been observed in the cases of galaxies collapsing into superclusters via filaments \citep[e.g][]{Brough2006,Porter2008,Mahajan2012}.

Recently, \citet{Alpaslan2015} concluded that stellar mass is the predominant factor of galaxy properties in the local Universe, as compared to gross environment (e.g. filament versus void; see Fig. 15 in \citealt{Alpaslan2015}, see also \citealt{Wijesinghe2012}). The present paper asks a complementary question: does the location of a galaxy relative to large scale filaments, the presumed channels of gas flow over cosmic time, predispose the galaxy to higher or lower stellar mass growth (either in the past or currently on-going in the form of star formation activity)? Specifically, we investigate what impact a galaxy's placement within a filament has on its stellar population, both integrated over its life (stellar mass) and the younger stellar component (UV-derived star formation rate), with the assumption that the properties and availability of gas vary with filament position. Our study is conducted exclusively on spiral galaxies free of morphological peculiarities, detected nuclear activity, and which are not a member of a galaxy group or pair. 

We describe our data in Section 2, from our large-scale structure catalogue and spiral galaxy selection to our UV-derived specific star formation rates. Section 3 focuses on our analysis, the results of which are subsequently discussed in Section 4. Section 5 provides a summary and conclusion. Throughout this paper, consistent with the cosmology used in \citet{Alpaslan2013a}, we adopt $\Omega_{\mathrm{m}} = 0.25,\; \Omega_{\Lambda}=0.75,\; H_0 = h\; 100 \mathrm{\;km\; s}^{-1}\;\mathrm{Mpc}^{-1}$.

\section{Data}

\subsection{GAMA and large-scale structure}

The characterisation of filaments is considerably challenging, due in large part to the complex morphologies of these structures. Observationally, studies of filaments require galaxy redshift surveys that are both highly complete, and have a high target density. One such survey is the Galaxy And Mass Assembly (GAMA, \citealp{Driver2009,Driver2011,Hopkins2013,Liske2015,Driver2016}) survey, which combines spectroscopic data obtained at the Anglo-Australian Telescope (AAT, NSW, Australia) with multiwavelength (\emph{UV-FIR}) photometric data from a number of ground and space based facilities. The spectroscopic campaign of the survey provides ~250,000 spectra for galaxies across 5 fields; $\alpha$ = 9h, $\delta$ = 0.5 deg (G09), $\alpha$ = 12h, $\delta$ = -0.5 deg (G12) and $\alpha$ = 14h30m, $\delta$ = 0.5 deg (G15), $\alpha$ = 2h, $\delta$ = -8.125 deg (G02) and $\alpha$ = 23h and $\delta$ = -32.5 deg (G23). The three equatorial fields (G09, G12 and G15) are $12 \times 5$ degrees each, and the two southern fields (G02 and G23) are respectively $8.6 \times 2.5$ and $12 \times 5$ degrees. The survey is $>98\%$ complete down to $m_r = 19.8$ mag. See \citet{Driver2016} for a detailed description of how all this data is homogenised and assimilated into a cohesive photometry catalogue, and \citet{Liske2015} for details on the spectroscopic component of the survey. The high target density and spectroscopic completeness of GAMA enables the present work to be a comprehensive search for correlations between large-scale structure and galaxy evolution.

A volume limited galaxy catalogue of 11,791 objects was constructed as a subset of the GAMA Large-Scale Structure Catalogue (GLSSC; \citealp{Alpaslan2013a,Alpaslan2014b}). This catalogue has an absolute magnitude limit of $M_r = -17.6 + 5 \log h$ mag out to $z = 0.09$; these limits are chosen such that the catalogue is complete to a stellar mass of $\geq 9 \log M_* / h^{-2} M_{\odot}$. Classification of galaxies as belonging to three types of large scale environments (filaments, tendrils, and voids) was performed on this catalogue via a modified minimal spanning tree (MST) algorithm \citep{Alpaslan2013a}. Galaxy groups from \citet{Robotham2011} are first used as nodes of a MST to find filaments; groups are considered to be in a filament if they are a distance $b$ from each other, similar to a friends-of-friends algorithm. Galaxies that are a distance $r$ from these filaments (or in a group in a filament) are defined as galaxies in filaments. All galaxies in filaments are then removed from the sample, and a second MST is generated on the remaining population. The second pass identifies tendrils, and any galaxy beyond a distance $q$ from a tendril is considered to be an isolated void galaxy. The distances $r$ and $q$ are chosen such that the two-point correlation function $\xi_2(r)$ of void galaxies is minimised. $b$ is chosen such that at least 90\% of groups with $L \geq 10^{11} L_{\odot}$ are in filaments. For other examples of MST-based structure finders, see \citet{Doroshkevich2004} and \citet{Colberg2007}. The GLSSC is generated with $b = 5.75 h^{-1}$ Mpc; $r = 4.12 h^{-1}$ Mpc; and $q = 4.56 h^{-1}$ Mpc. The catalogue used for this work is generated with $b = 3.7 h^{-1}$ Mpc; $r = 3.79 h^{-1}$ Mpc; and $q = 4.35 h^{-1}$ Mpc. 

An alternative filament finding algorithm based around the tidal tensor prescription has recently been used on GAMA data by \citet{Eardley2015}. The tidal tensor methodology differs significantly from the MST approach used to generate the GLSSC, in that it is a density-based method that categorises volumes as belonging to knots, sheets, filaments, or voids. \citet{Eardley2015} present a comparison of their galaxy classifications to the GLSSC in their appendix, and find that the two catalogues are in excellent agreement, particularly when it comes to categorising galaxies as being in filaments. Such an agreement indicates that both methodologies are able to identify filamentary structures consistently: 92.6\% of GLSSC filament galaxies are in sheets, filaments, and knots defined by the tidal tensor method, with the majority (44.3\%) being filament-filament matches. For further details we refer the reader to Appendix C and Figures C1 and C2 of \citet{Eardley2015}. A more comprehensive review and comparison of how a number of filament finding methods, including the MST algorithm used for the GLSSC, perform when run on the same data set will be presented by Libeskind et al. (in prep).

\subsection{Spiral galaxy selection}

As mentioned in the introduction, this work focuses exclusively on spiral galaxies that are in filaments. We vet against spheroidal galaxies, as their stellar population is influenced by a different set of dynamics within filaments \citep{Tempel2014}, as well as by local dynamical effects such as tidal forces and galaxy interactions, influenced by halo mass; these are all known to directly influence the SFR of a galaxy \citep{Robotham2013a}. Concentrating on spiral galaxies, whose SFRs are related to the properties of their surrounding intergalactic medium (Grootes et al. in prep), and selecting those which do not belong to groups or pairs, provides us the means to investigate whether galaxies have actively changing stellar populations influenced by their position along a filament.

Formally, this criterion means a selection of spiral galaxies that are within $r = 3.79 h^{-1}$ Mpc from a filament, consistent with the maximum allowed distance between a galaxy and a filament as described in the preceding subsection. Furthermore, we only consider spiral galaxies that meet both this distance criterion and are not members of groups or galaxy pairs; this ensures a selection of isolated galaxies without satellites; this constitutes about 25\% of all spirals in filaments. Group membership information is taken from the \citet{Robotham2011} catalogue, and a galaxy is considered to be in a pair if it has a companion within a physical projected separation of 100 h$^{-1}$ kpc and a velocity separation of 1000 kms$^{-1}$. Our isolated galaxies are therefore considered to be isolated down to GAMA's detection limit of $m_r = 19.8$ mag. The accuracy of these group and pair classifications has been indepently verified \citep{Alpaslan2012}, and benefits from GAMA's high target density and spectroscopic completeness, ensuring that our isolation criteria for our sample is very robust. By excluding such paired and grouped galaxies, one can study the stellar mass growth of spiral galaxies whose stellar populations and star formation rates are not affected by processes that typically affect star formation in groups and pairs \citep[e.g][]{Ellison2008,Robotham2013a,Davies2015}.

\begin{figure}
	\includegraphics[width=1\columnwidth]{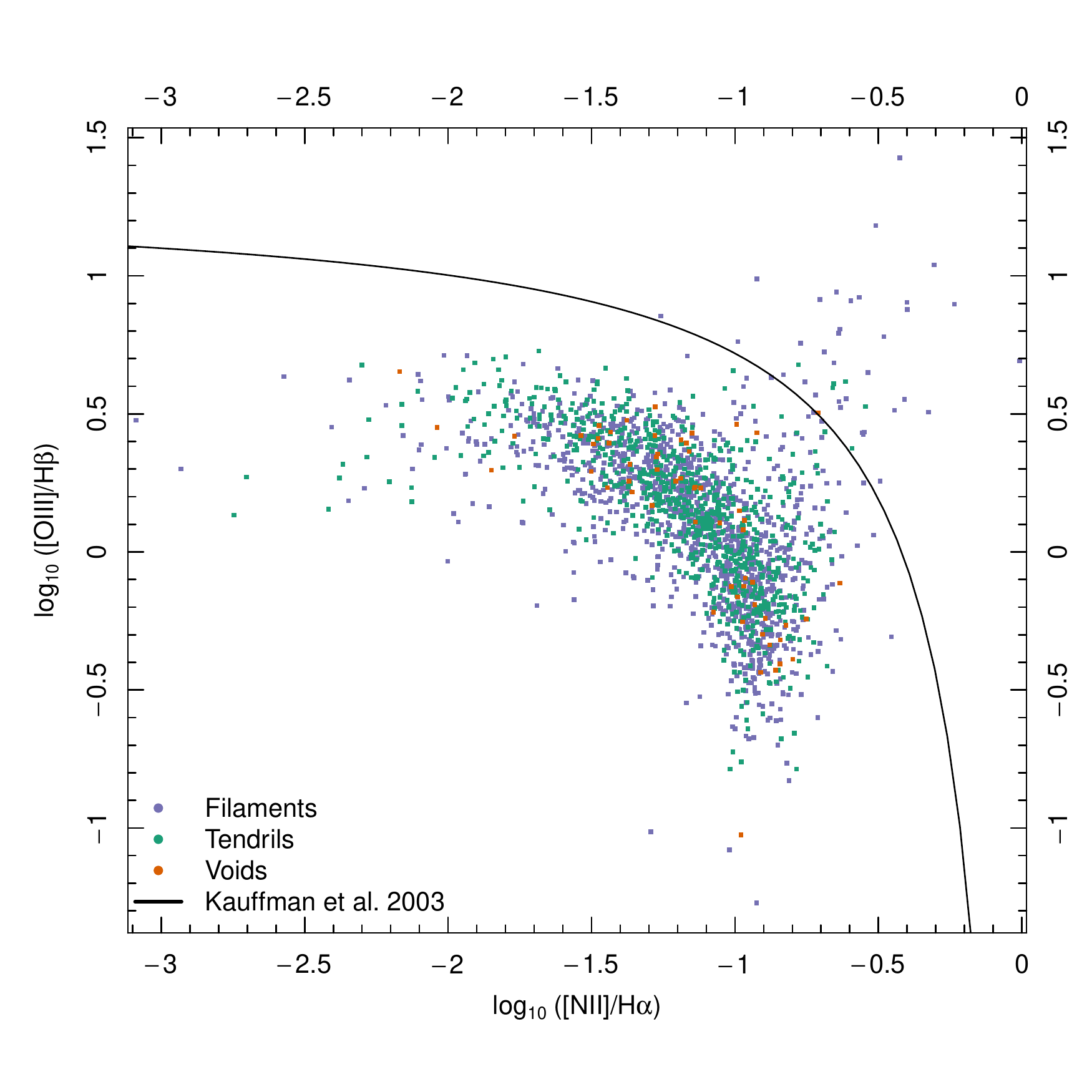}
	\caption{BPT diagram displaying the \citet{Kauffmann2003} relation, distinguishing star forming galaxies from AGN/composite objects, for spiral galaxies in filaments, tendrils, and voids (blue, green, and orange respectively). We reject all galaxies lying above the line as non-star formers.}
	\label{fig:BPT}
\end{figure}

\begin{figure*}
	\includegraphics[width=\textwidth]{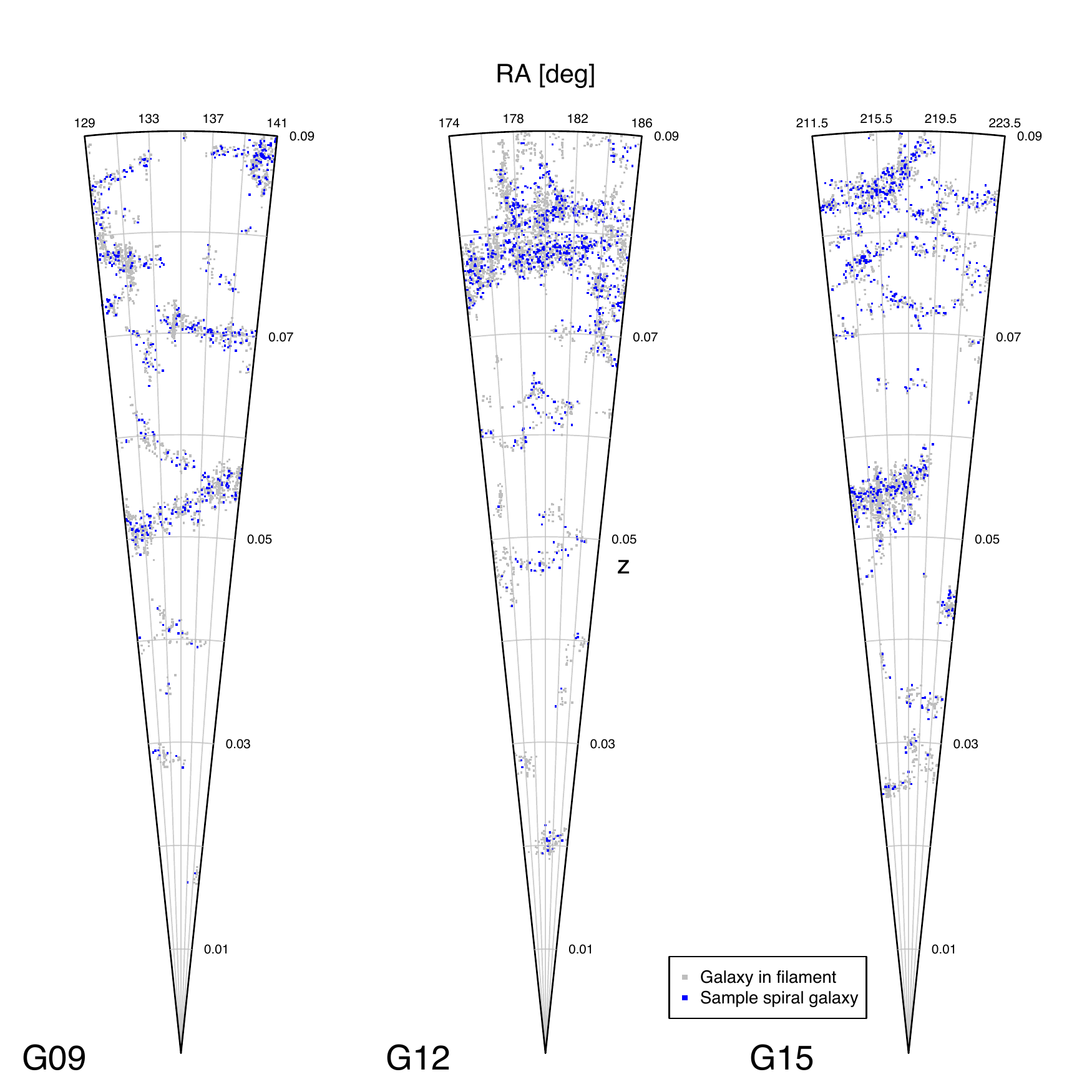}
	\caption{Overhead view of the three equatorial GAMA fields, with all filament galaxies in the low redshift large-scale structure catalogue shown as grey points. Of those, non-AGN spiral galaxies that are not in groups or pairs are shown as blue points.}
	\label{fig:cones}
\end{figure*}

We identify spiral galaxies using the non-parametric, cell-based methodology described in \citet{Grootes2013}, where combinations of two or three photometric parameters are used to divide populations of galaxies into morphological type. Morphological classifications from Galaxy Zoo Data Release 1 \citep{Lintott2011} are used to calibrate this division. For each parameter combination, the space occupied by spiral galaxies identified from the Galaxy Zoo data set is divided into a series of cells whose sizes depend on local density of points. A given combination of parameters is judged to be successful at identifying spiral galaxies if a sufficient fraction of galaxies in these cells are indeed galaxies identified as spirals from Galaxy Zoo. A number of parameter combinations are found to perform well, and for this work we utilise the combination of $\log(n)$, $\log(r_e)$, and $M_i$, where $n$, $r_e$ and $M_i$ are the S\'{e}rsic index, effective radius, and $i$-band absolute magnitude of each galaxy. S\'{e}rsic indices and  effective radii for each galaxy are taken from \citet{Kelvin2012}. We additionally reject 45 non star-forming galaxies as per the \citet{Kauffmann2003} BPT relation, displayed in Fig. \ref{fig:BPT}. Line strengths in Fig. \ref{fig:BPT} are directly measured from GAMA spectra in a manner similar to that described in \citet{Gunawardhana2013} using \textsc{mpfitfun}\footnote{http://www.physics.wisc.edu/~craigm/idl/fitting.html}, but with additional error modeling for each spectral fit. Our final sample contains 1799 star forming spiral galaxies across the three equatorial GAMA fields, whose morphologies have been verified by independent visual inspection. An overhead view of these galaxies, as well as the overall filamentary structure that they reside in is shown in Fig. \ref{fig:cones}.

\subsection{UV derived star formation rate measurements}

The SFR measurements provided in this paper are based on UV emission, corrected for Galactic and internal dust attenuation. Not only does the NUV provide an estimate of the SFR of a galaxy on the timescale of $\sim 10^8\,$yr \footnote{For a galaxy with a constant SFR the restframe luminosity-weighted mean age of the GALEX NUV filter is $\sim10^8\,$yr (\citealp{Gilbank2010}, Grootes et al. in prep).}, the GAMA NUV measurements also provide robust estimates of the total NUV flux, and accordingly the total SFR.

Coverage of the GAMA fields in the NUV is provided by GALEX in the context of GALEX MIS \citep{MARTIN2005,MORRISSEY2007} and by a dedicated guest investigator program (\textit{GALEX-GAMA}), providing a largely homogeneous coverage to $\sim23\,\mathrm{mag}$. Details of the GAMA UV photometry are provided in Andrae et al. (in prep), \citet{Liske2015} and on the GALEX-GAMA website\footnote{www.mpi-hd.mpg.de/galex-gama/}. Briefly however, extraction of UV photometry proceeds as follows: GAMA provides a total of three measurements of NUV fluxes. Firstly, all GALEX data is processesd using the GALEX pipeline v7 to obtain a uniform blind source catalogue \footnote{The band merged GALEX blind catalogue is $NUV$-centric, i.e. $FUV$ fluxes have been extracted in $NUV$ defined apertures, entailing that no catalogued source can be detected only in the $FUV$} with a signal to noise ($S/N$) cut at $2.5\,\sigma$. This catalogue has subsequently been matched to the GAMA optical catalogue using an advanced matching technique which accounts for the possibility of multiple matches between optical and NUV sources, redistributing flux between the matches as described in Andrae et al. (in prep.) and on the GALEX-GAMA website. Additionally NUV photometry at the positions of all GAMA target galaxies is extracted using a curve-of-growth algorithm, as well as apertures defined on the measured size of the source in the $r$-band. For one-to-one matches preference is given to the pipeline photometry, while for extended sources and multiple matches, the curve-of-growth and aperture photometry is preferred. The resulting best estimates of the total NUV flux of the galaxy is reported as \texttt{BEST\_FLUX\_NUV}, in the UV photometric catalogue. In the work presented here, we have made use of these estimates applying galactic foreground extinction corrections following\citep{Schlegel1998}\footnote{In the NUV we make use of $A_{\mathrm{NUV}} = 8.2\,E(B-V)$ as provided by \citet{WYDER2007} }, and k-corrections to $z=0$ using \texttt{kcorrect\_v4.2} \citep{BLANTON2007}. 

The determination of the SFR of a galaxy from its NUV emission requires \textit{intrinsic} emission which has been corrected for the attenuation of the stellar emission due to the dust in the galaxy, which is particularly severe at short (UV) wavelengths \citep[e.g.][]{Tuffs2004}. Our analysis is focussed on the differential effects in the SFR of spiral galaxies as a function of (large-scale) environment. However, both the so-called `main sequence of star forming galaxies' \citep[e.g.][]{NOESKE2007,WHITAKER2012} as well as work on the specific star formation rate - stellar mass relation for purely morphological samples of spiral galaxies (\citealp{GROOTES2014}, Grootes et al. 2015 in prep) imply that environmental effects are likely of second order and superimposed on the dominant influence of galaxy properties such as stellar mass. Accordingly, we require a method of obtaining accurate attenuation corrections which is as free as possible of both systematic and random errors.

In this paper we have adopted the method of \citet{Grootes2013} which uses the radiation transfer model of \citet{POPESCU2011}, and supplies attenuation corrections on an object-by-object basis for spiral galaxies, taking into account the orientation of the galaxy in question. As demonstrated by \citet{Grootes2013} the optical depth due to dust, critically determining the attenuation of emission from a galaxy, can be estimated using the effective stellar mass surface density $\mu_*$, thus enabling the determination of highly accurate attenuation corrections for large samples of galaxies on an object-by-object basis. In determining attenaution corrections we have proceeded as follows:

The GAMA measurements of galaxy stellar mass and size have been used to determine the effective stellar mass surface density $\mu_*$ as

\begin{equation}
\mu_* = \frac{M_*}{2 \pi D^{2}_{A}(z)\theta_{e,ss,r}^{2}}\;,
\end{equation}
where $D_{A}(z)$ is the angular diameter distance corresponding to the redshift $z$, $M_*$ is the stellar mass, and $\theta_{e,ss,r}$ is the angular size corresponding to the effective radius of the $r$-band single S\'ersic profile. Subsequently, $\mu_*$ is used to estimate the optical depth, which combinded with a measurement of the inclination of the galaxy, is used to predict the attenuation in the UV-optical bands using the model of \citet{POPESCU2011}. The reader is referred to \citet{Grootes2013} for further details of the process and the \citet{POPESCU2011} model.

Using the intrinsic absolute foreground extinction corrected NUV magnitudes derived in this manner we estimate the SFR using the conversion given in \citet{KENNICUTT1998} scaled from a \citet{SALPETER1955} IMF to a \citet{CHABRIER2003} IMF as in \citet{SALIM2007}, i.e.

\begin{equation}
SFR [M_{\odot} \mathrm{yr}^{-1}] = \frac{L_{\mathrm{NUV}} [\mathrm{Js}^{-1}\mathrm{Hz}^{-1}]}{1.58\times 7.14\cdot 10^{20}}
\label{eq_SFR}
\end{equation}

\begin{figure}
	\centering
	\includegraphics[width=0.5\textwidth]{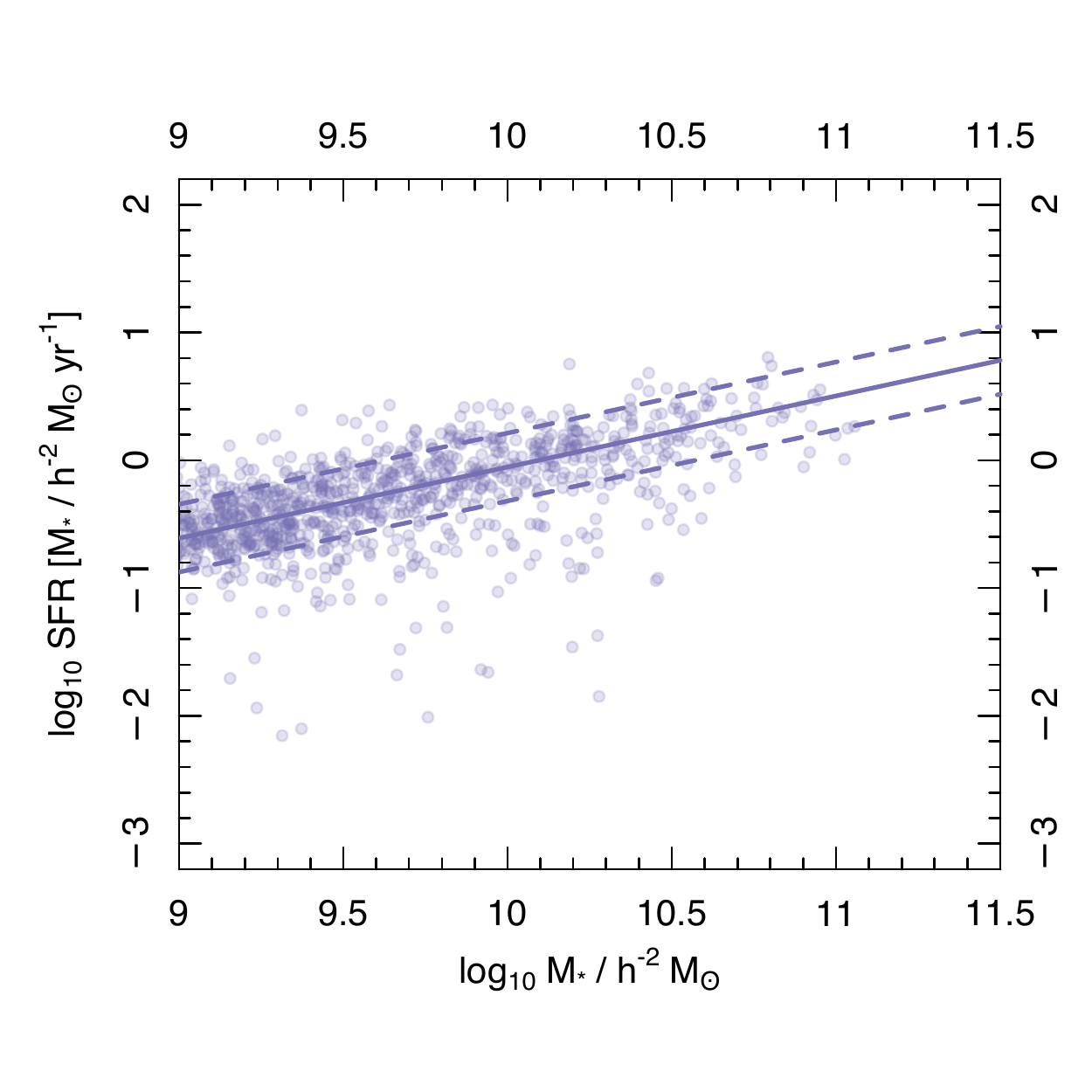}
	\caption{Star formation rate as a function of stellar mass for ungrouped, unpaired, non-AGN spiral galaxies in filaments. The model fit to the data is shown as the solid line, with $1\sigma$ contours shown as the dashed lines above and below the median.}
	\label{fig:massvSFR_fils}
\end{figure}

\noindent The SSFR $\psi_*$, the specific star formation rate, is computed by dividing the SFR for each galaxy by its stellar mass $M_*$. GAMA stellar masses are calculated from synthetic spectra fit to rest frame $ugriz$ photometry for each galaxy, see \citet{Taylor2011} for further details. For two galaxies in our sample this process yields unphysically high SFR estimates due to poor size estimates. These errors lead to an overestimation of their dust content, and consequently an overattenuation of their UV flux. These two galaxies are ommitted from the subsequent analysis in this paper. We show the relationship between stellar mass and star formation rate\footnote{The line fitting in all figures is done using the \textsc{hyperfit} package (\citealp{Robotham2015}, http://hyperfit.icrar.org). \textsc{hyperfit} uses either downhill searches or MCMC (Markov chain Monte Carlo) methods to calculate the best-fit parameters for a hyperplane of $D-1$ dimensions for $D$ dimensional data. For this work we run \textsc{hyperfit} using its Nelder-Mead downhill simplex \citep{Nelder1965} implementation.} for these selected filament galaxies in Fig. \ref{fig:massvSFR_fils}.

\subsection{Distance metrics}
\label{sec:metrics}

Filaments, despite their relative linearity, are difficult structures to characterise geometrically, and this difficulty is further enhanced by the fact that filaments are made up of discrete particles (galaxies) as opposed to continuous distributions of matter. Further, objectively defining the `centre' of a filament and the distance to that centre from a given galaxy are non-trivial tasks. For recent discussions on this particular issue, we refer the reader to \citet{Aragon-Calvo2010} and \citet{Cautun2014}.

We build upon the work of structurally decomposing filaments that was presented in \citet{Alpaslan2013a}, specifically in Appendix A. There, the centre of a filament is defined as the group that is furthest away from any of the edges of the filaments, within the structure imposed on it by the minimal spanning tree formed between its constituent groups. `Furthest away' can be defined both in terms of physical 3D comoving distance between the groups, as well as simply the number of groups between the centre and the edge. In this work, the former definition is used. The longest continuous path of links from one end of the filament to another through the filament centre is defined as the first order branch or backbone, and represents the primary `axis' of the filament. Paths that lead into the backbone are referred to as second order branches. Generally, paths that lead into $n^{\mathrm{th}}$ order branches are referred to as $(n+1)^{\mathrm{th}}$ order branches.

\begin{figure}
	\centering
	\includegraphics[width=0.5\textwidth]{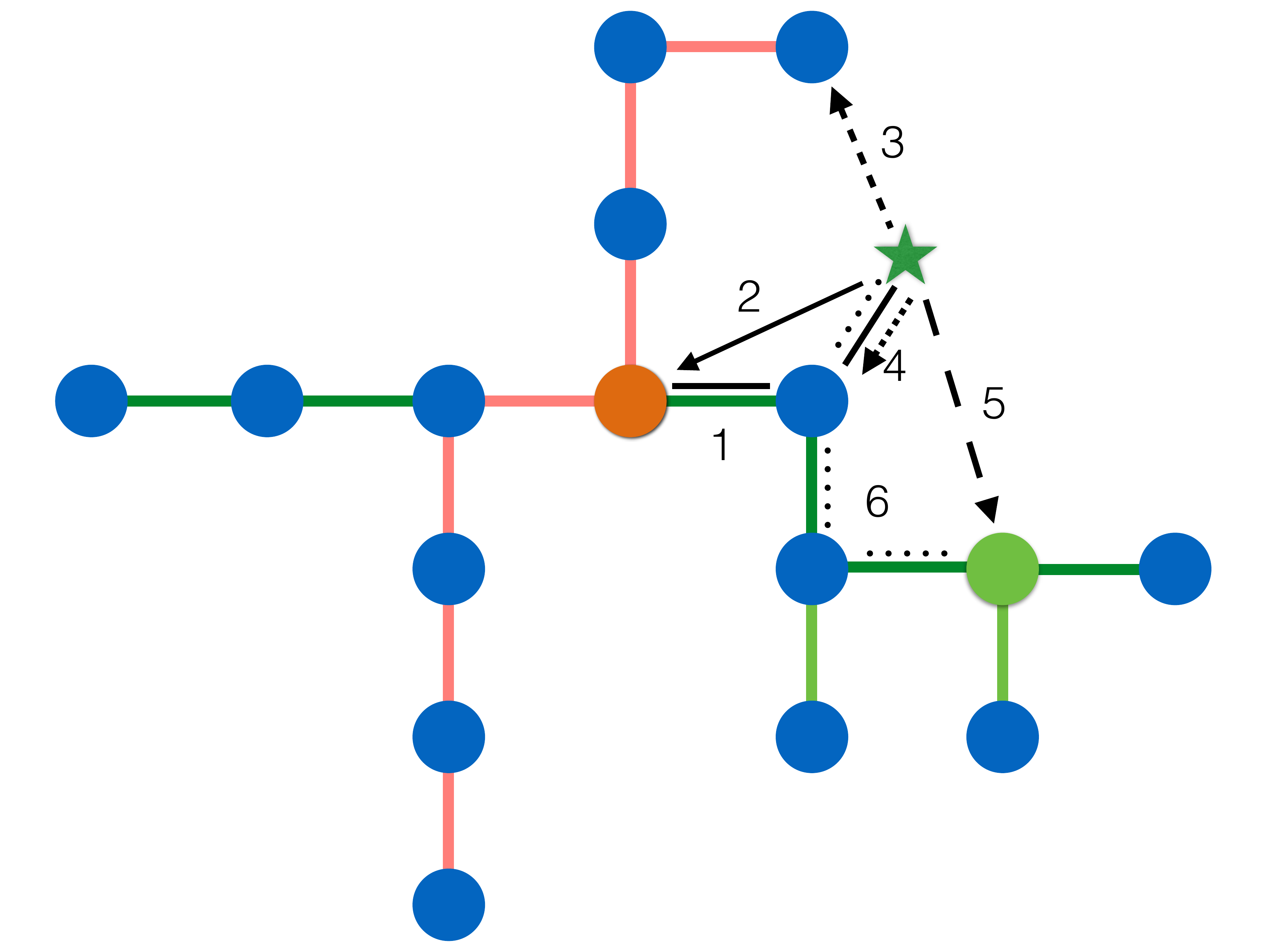}
	\caption{Schematic of a filament, with its constituent groups represented as large blue circles. The orange and green circles represent the central and most massive groups, and the star represents a hypothetical spiral galaxy within the filament. Links are coloured according to their branch order, with the backbone shown as the thick pink links. Second and third order branches are dark and light green respectively. The six distance metrics we use are shown as different lines, and are numbered according to their description in S\ref{sec:metrics}.}
	\label{fig:schematic}
\end{figure} 

\begin{figure*}
\includemovie[
	3Dviews2 = views.tex,
	poster,
	toolbar,
	label=filament3d.u3d,
	text={\includegraphics[width=\textwidth]{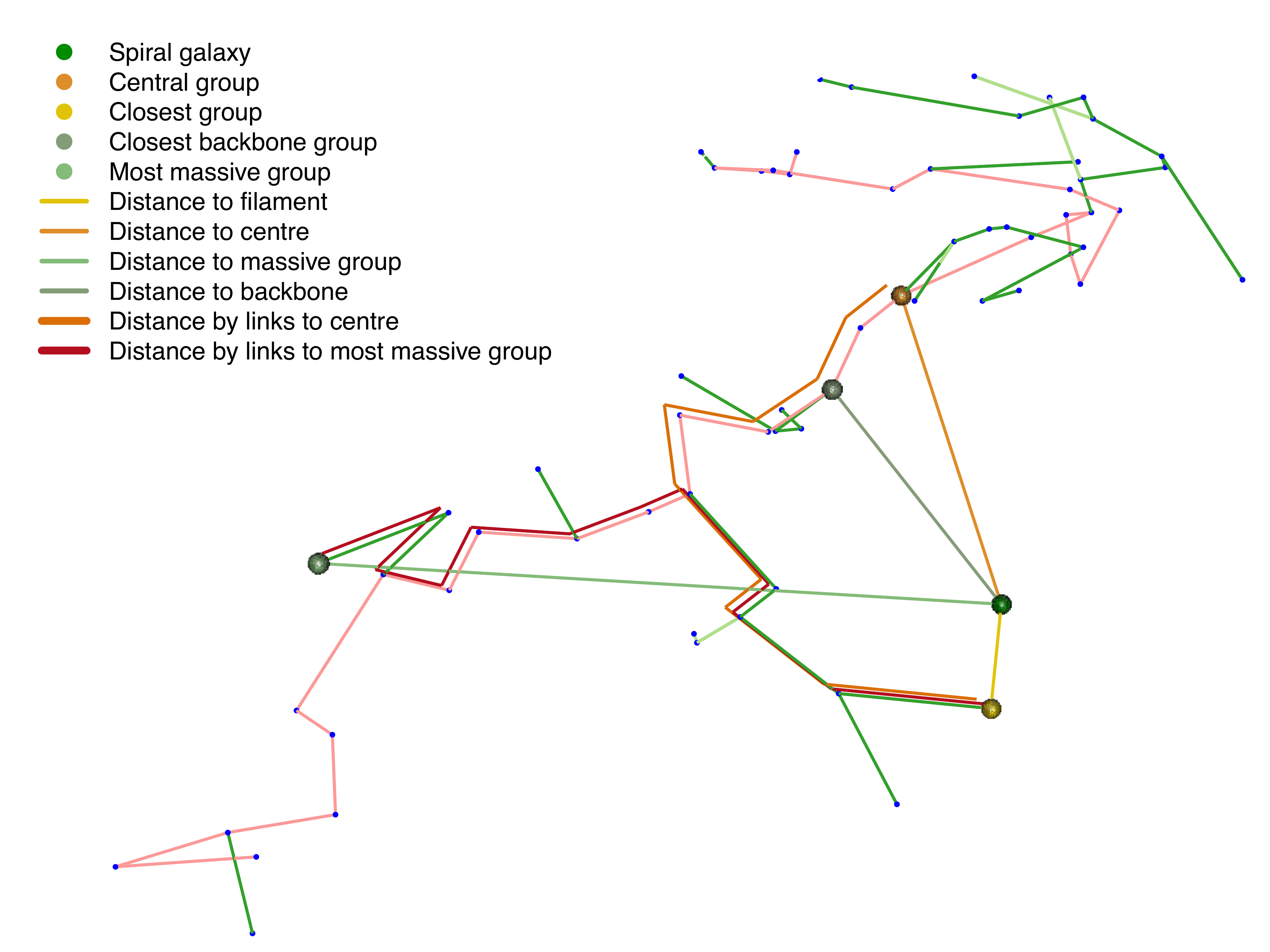}},
	3Daac=60.000001669652114, 3Droll=-85.13789136266956, 3Dc2c=0.05373663082718849 -0.612947940826416 -0.7882938385009766, 3Droo=26.181367215212664, 3Dcoo=-18.245609283447266 2.6941614151000977 225.68634033203125,
	3Dlights=CAD,
]{\linewidth}{\linewidth}{filament3d.u3d}
	\caption{A 3D, interactive model of a filament, and a spiral galaxy within it. If the Figure does not appear interactive, please view with a current version of Adobe Reader. The filament is represented as a series of links that connect its constituent groups, which are shown as blue cubes. The links are coloured according to their rank, i.e. the backbone links are shown as pink lines, and second and third order branches are shown as green and light green lines. The spiral galaxy, central group, closest group, closest backbone group and most massive group are shown as green, orange, yellow, grey, and teal; and the lines representing the distances to those groups are shown in matching colours. Finally, the distance by links to the central group and most massive group are shown as red and burgundy lines, offset from the filament links. Views for the interactive figure showing all of these components may be accessed from the toolbar at the top of the interactive figure, and the 3D figure can be viewed as a separate window by selecting the relevant option from the contextual menu.}
	\label{fig:fil3d}
\end{figure*}

These considerations lead to 6 different ways to characterise the distance between spiral galaxies and the filaments that they reside in. The distance metrics fall into two broad groups: (1) direct radial distances between the spiral galaxy and different components of the filaments, e.g. the central group, and (2) distance traveling along the filament links from the spiral galaxy to different filament components. These two categories probe slightly different aspects of gas flow within filaments: the first is more sensitive to radial changes in the graviational potential well of a filament and how that influences star forming gas, while the second probes the transversal changes in gas properties as it travels along the links within a filament. Figure \ref{fig:schematic} provides an overview of the metrics that we use in this work for an idealised two dimensional case. An interactive 3D model of a filament with these same distance metrics is shown in Fig. \ref{fig:fil3d}. 
\begin{enumerate}
\item the distance from the target galaxy to the central group of the filament, along the links in the filament. The radial distance from the galaxy to the nearest filament group is added to this. The central group is defined as the group that is furthest from any edge of the filament (see \citealp{Alpaslan2013a}); 
\item the radial distance from the target galaxy to the central group of the filament;
\item the radial distance from the target galaxy to the nearest group in the backbone of the filament; 
\item the orthogonal distance from the target galaxy to the filament\footnote{This distance metric is originally defined in \citet{Alpaslan2013a}, where it was used to determine if a galaxy was close enough to a filament to belong to it. For a given galaxy, the distance to the filament is defined as the radial distance between the galaxy and the nearest filament group or filament link, depending on which is shortest.};
\item the radial distance from the target galaxy to the most massive group in the filament, where mass is defined as the sum of the stellar mass of its constituent galaxies; and
\item distance by links from the target galaxy to the most massive group in the filament. As with the first metric, the radial distance to from the galaxy to the nearest filament group is also added to this.
\end{enumerate}


\section{Stellar content along the cosmic web}

\subsection{Filaments}
\label{sec:filoffset}

For each galaxy in our sample, we compute its distance relative to its filament according to the 6 metrics described in the preceding section. To avoid recovering star formation trends driven by stellar mass instead of filament morphology, we compute a SSFR offset value for each galaxy, similar to Grootes et al. (in prep). We begin by calculating median values of SSFR ($\psi$) in bins of stellar mass for the sample shown in Fig. \ref{fig:massvSFR_fils}, and interpolating between them. The interpolated relation is fit by the line $y = -0.35x - 6.54$. Each galaxy's SSFR offset is then defined as $\psi_{\mathrm{galaxy}} - \psi_{\mathrm{med}}$, where $\psi_{\mathrm{med}}$ is a median SSFR value calculated for the stellar mass of that particular galaxy from our interpolated medians; errors in star formation are included in each fit.

\begin{figure}
	\centering
	\includegraphics[width=\columnwidth]{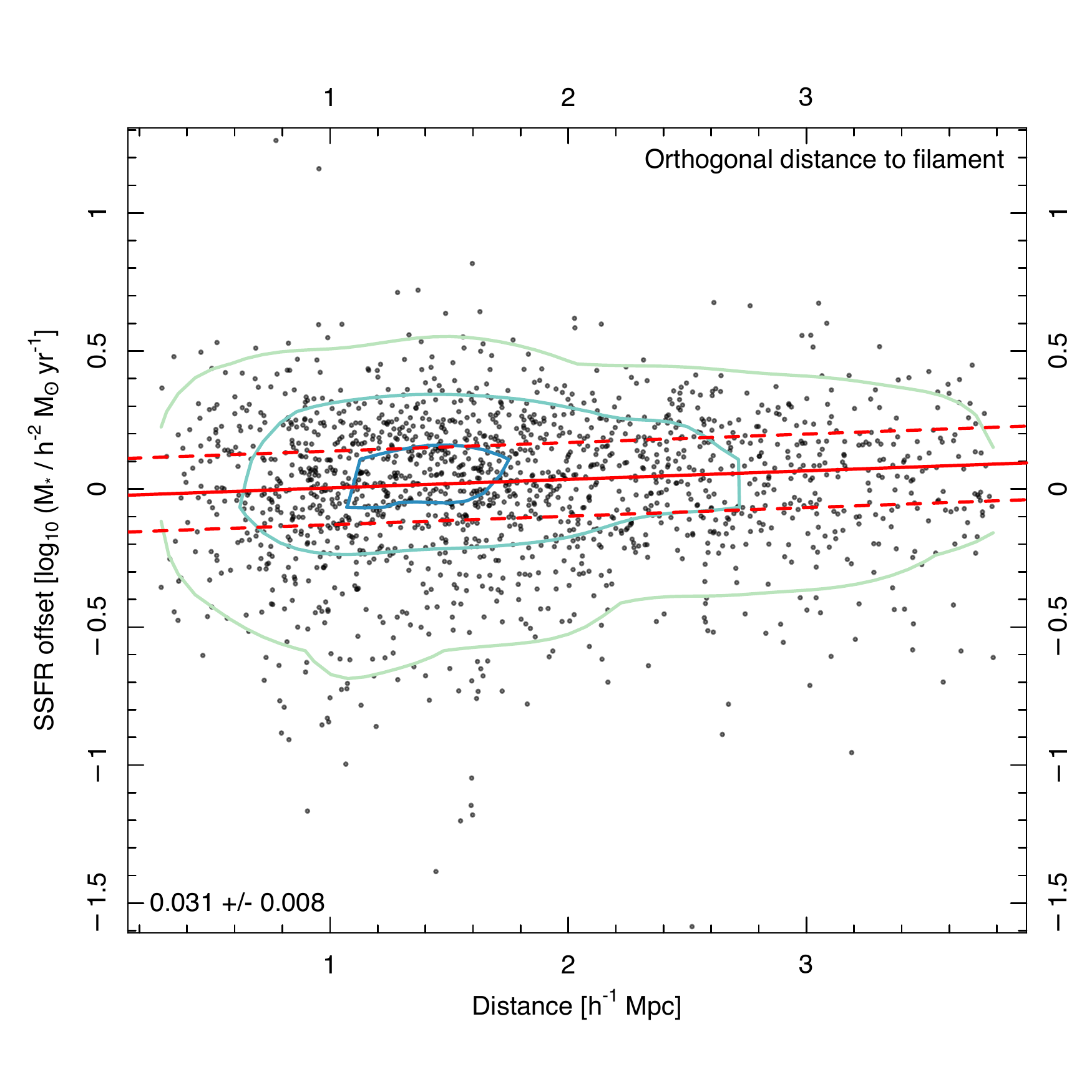}
	\caption{Offset in UV-derived specific star formation rates of spiral galaxies in filaments as a function of their orthogonal distance to their host filament. The solid lines show fits to the data using the \textsc{hyper.plane} fitting package \citet{Robotham2015}, and the dashed lines represent $1\sigma$ errors on the fit. Coloured contours show the 90$^{\mathrm{th}}$, 50$^{\mathrm{th}}$, and 10$^{\mathrm{th}}$ percentiles of data respectively. The gradient and associated error for the fit is shown in the bottom left. The fit indicates that the star formation rates of spiral galaxies in filaments are higher closer to the edge of the filament when compared to the core of the filament. The detection of such an effect in ungrouped and unpaired spiral galaxies means that the most likely explanation for this change in star formation rates is a change in the availability of gas to these galaxies as a function of their orthogonal distance to the filament core.}
	\label{fig:dist_SFR_distToFil}
\end{figure}

We perform linear fits to the data for each metric, and find that in most cases, the errors are consistent with the slope of the fit being zero. To ensure that a linear fit is most appropriate, we compute Bayesian information criteria \citep{schwarz1978} for linear and polynomial fits and find that the BIC for a linear model is lowest by a margin of $\Delta{\mathrm{BIC}} > 10$. For the distance metric that measures the orthogonal distance between the galaxy and the filament we detect a very faint upward slope of $0.031 \pm 0.008$, which is shown in Figure \ref{fig:dist_SFR_distToFil}. The remaining fits are included at the end of this paper, in Figure \ref{fig:dist_SFR_all}. In both figures, the fit is shown as the solid line, with $1\sigma$ errors in the fit represented by the dashed lines. The trend in this Figure is statistically significant at a $3\sigma$ level and is consistent with a very subtle enhancement of specific star formation rates for galaxies closer to voids (i.e, on the outer edge of a filament) compared to those near the cores of filaments. A number of different scenarios may explain this relationship: galaxies on the peripheries of filaments are more exposed to gas, which is subsequently used to fuel new star formation; gas at the core of a filament is hotter than at the peripheries, and therefore more difficult to accrete; or galaxies traveling from the periphery of a filament to its core may simply be using up their gas reservoirs. These scenarios are all explored in greater detail in the discussion section. As a way of checking if signals in other distance metrics are washed out at long distances, we have repeated this calculation and considered only galaxies within 5 h$^{-1}$ Mpc of the filament for each distance metric; as well as fitting jackknife resampled distributions (where we reject a random 10\% of galaxies for each iteration), and our results remain unchanged.

\begin{figure}
	\centering
	\includegraphics[width=\columnwidth]{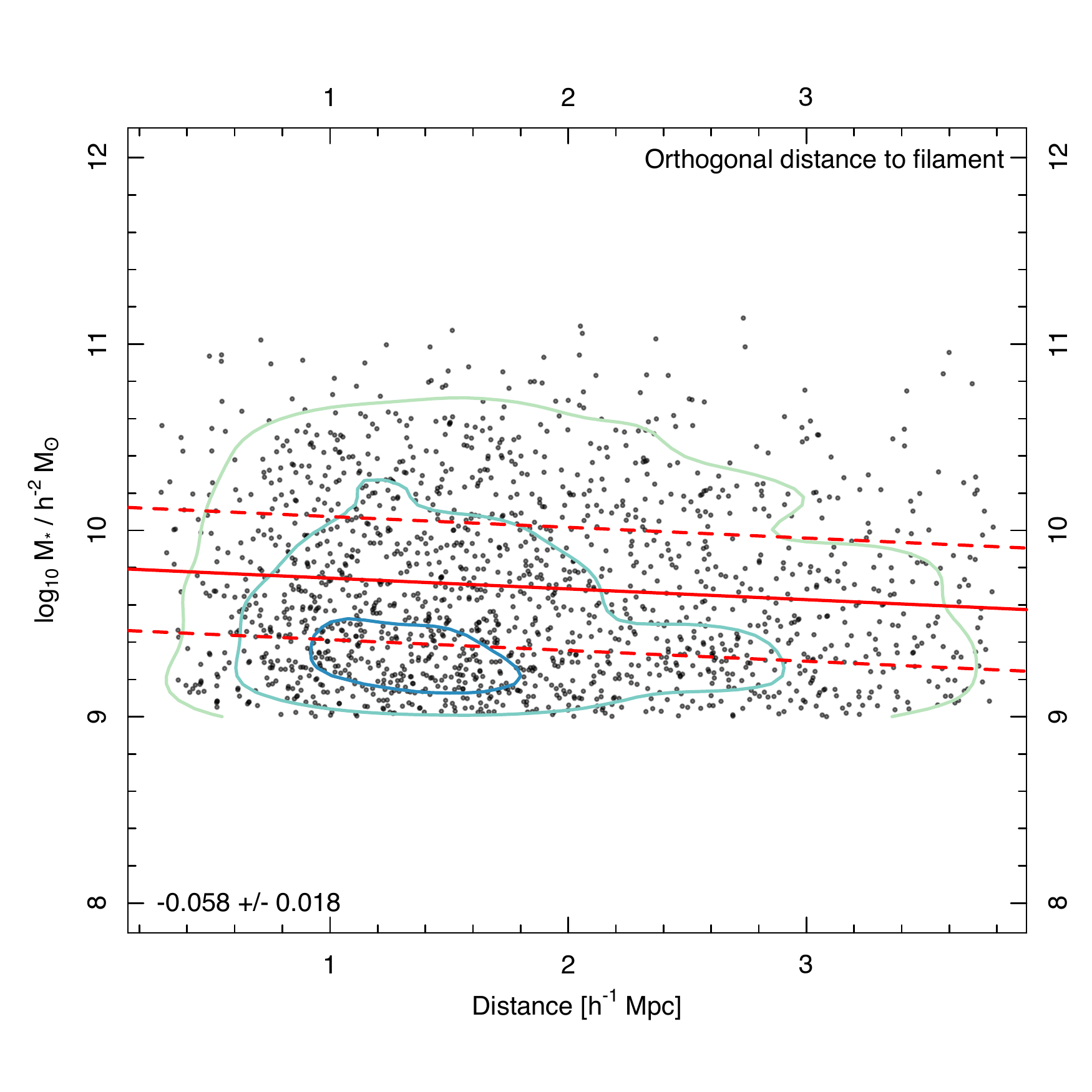}
	\caption{Stellar masses of spiral galaxies in filaments as a function of orthogonal distance to their host filament. The line fits in this Figure are done in the same way as those in \ref{fig:dist_SFR_distToFil}. Coloured contours show the 90$^{\mathrm{th}}$, 50$^{\mathrm{th}}$, and 10$^{\mathrm{th}}$ percentiles of data respectively. We detect a statistically significant downward trend in stellar mass as a function of orthogonal distance to the filament, showing that the stellar mass of spiral galaxies rises towards the core of a filament, and drops at its outer edges.}
	\label{fig:dist_mass_distToFil}
\end{figure}

We also track stellar mass in these galaxies as a function of our distance metrics. Once again there are no statistically significant trends in stellar mass as a function of most of the distance metrics except for the orthogonal distance to the filament, which is shown in Fig. \ref{fig:dist_mass_distToFil}, while the full set of fits is at the end of the paper, in Fig. \ref{fig:dist_mass_all}. Here there is a statistically significant slope of $-0.058 \pm 0.018$, indicating that galaxies closer to the edges of filaments have lower stellar masses than those in the cores; this is in good agreement with the recent work of \citet{Chen2015}. Given that our sample consists of galaxies outside of groups and pairs, this rise in stellar mass is unlikely due to be caused by local environmental effects (i.e. galaxy-galaxy interactions within groups and pairs). Coupled with the trend seen for the same distance metric in Fig. \ref{fig:dist_SFR_distToFil}, these two figures indicate that galaxies on the peripheries of filaments are forming stars more efficiently despite having lower masses, while those at the cores of filaments are heavier and forming fewer stars.

\begin{figure}
	\centering
	\includegraphics[width=\columnwidth]{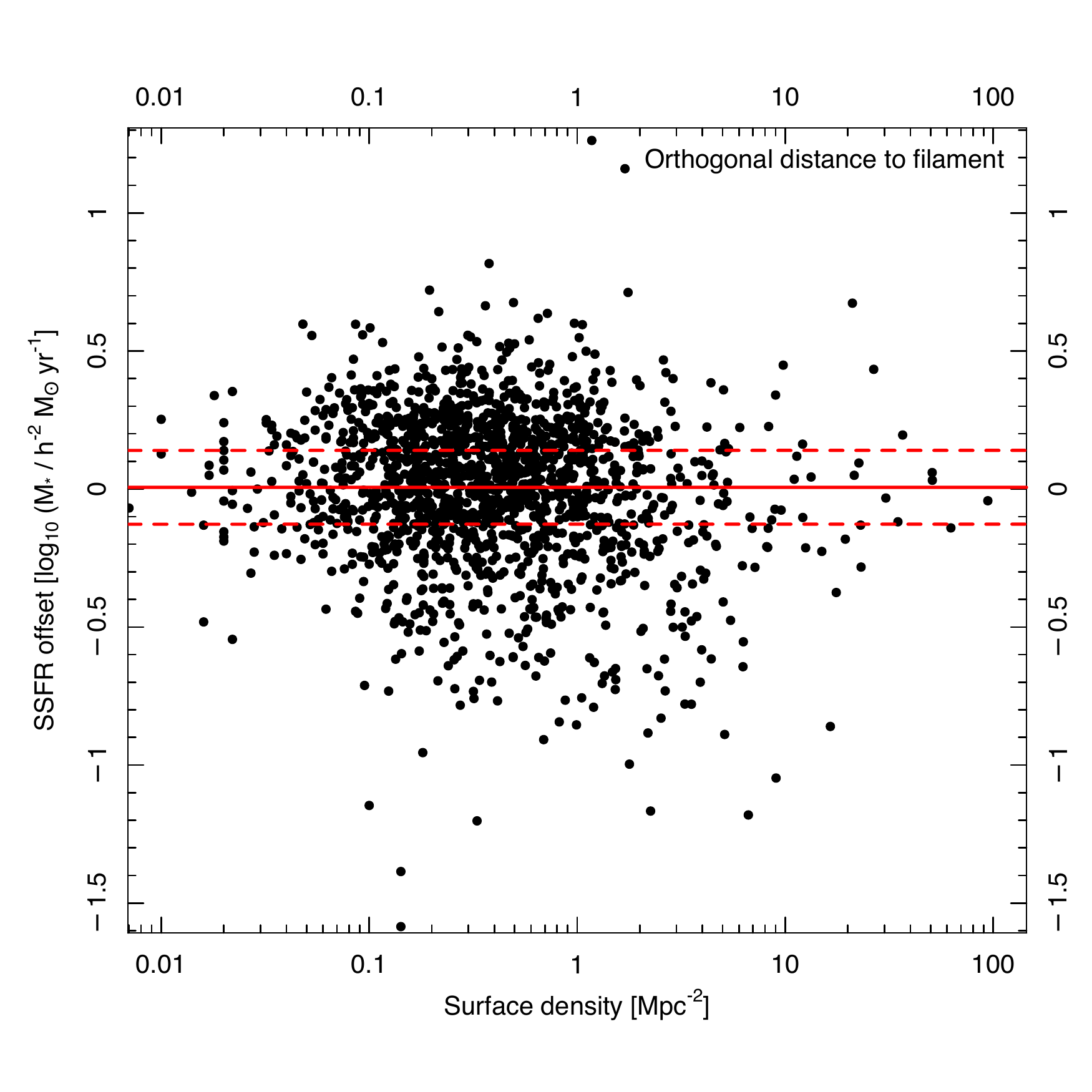}
	\caption{Surface density as defined by the 5 nearest neighbours of each galaxy from \citet{Brough2013} as a function of SSFR offset for filament spiral galaxies. We detect no statistically signficant correlation between these two parameters.}
	\label{fig:densitySSFR}
\end{figure}

\begin{figure}
	\centering
	\includegraphics[width=\columnwidth]{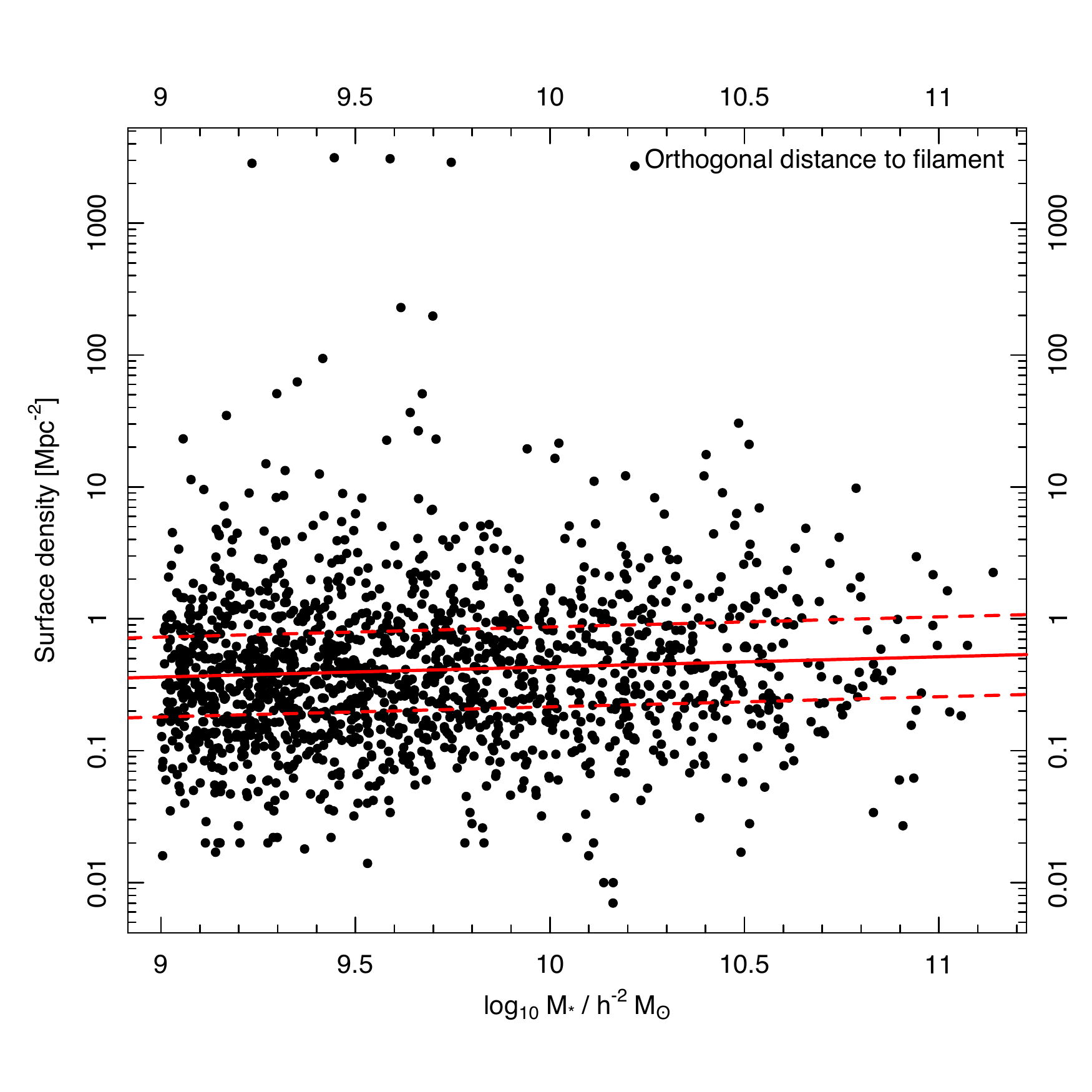}
	\caption{Stellar mass as a function of surface density as defined by the 5 nearest neighbours of each galaxy from \citet{Brough2013} for filament spiral galaxies. We detect no statistically signficant correlation between these two parameters.}
	\label{fig:densityMass}
\end{figure}

We check against the possibility that these effects are simply caused by the local density of our sample galaxies by plotting the SSFR offset and stellar masses of galaxies as a function of their local density, using the \texttt{EnvironmentMeasuresv03} catalogue of \citet{Brough2013}, where local densities are calculated for all GAMA galaxies. The surface density of each galaxy is calculated based on the distance to its $5^{\mathrm{th}}$ nearest neighbour within a velocity cylinder of $\pm 1000 \,\mathrm{km}\,\mathrm{s}^{-1}$, and are corrected for survey completeness. These are shown in Figs. \ref{fig:densitySSFR} and \ref{fig:densityMass}. We find that neither the SSFR offset, or the stellar mass of our sample galaxies depend at a statistically signficant level on their surface densities, providing further indication that the effects seen in Figs. \ref{fig:dist_SFR_distToFil} and \ref{fig:dist_mass_distToFil} are driven by large-scale structure.

\subsection{Comparison to tendrils and voids}

\begin{figure}
	\centering
	\includegraphics[width=\columnwidth]{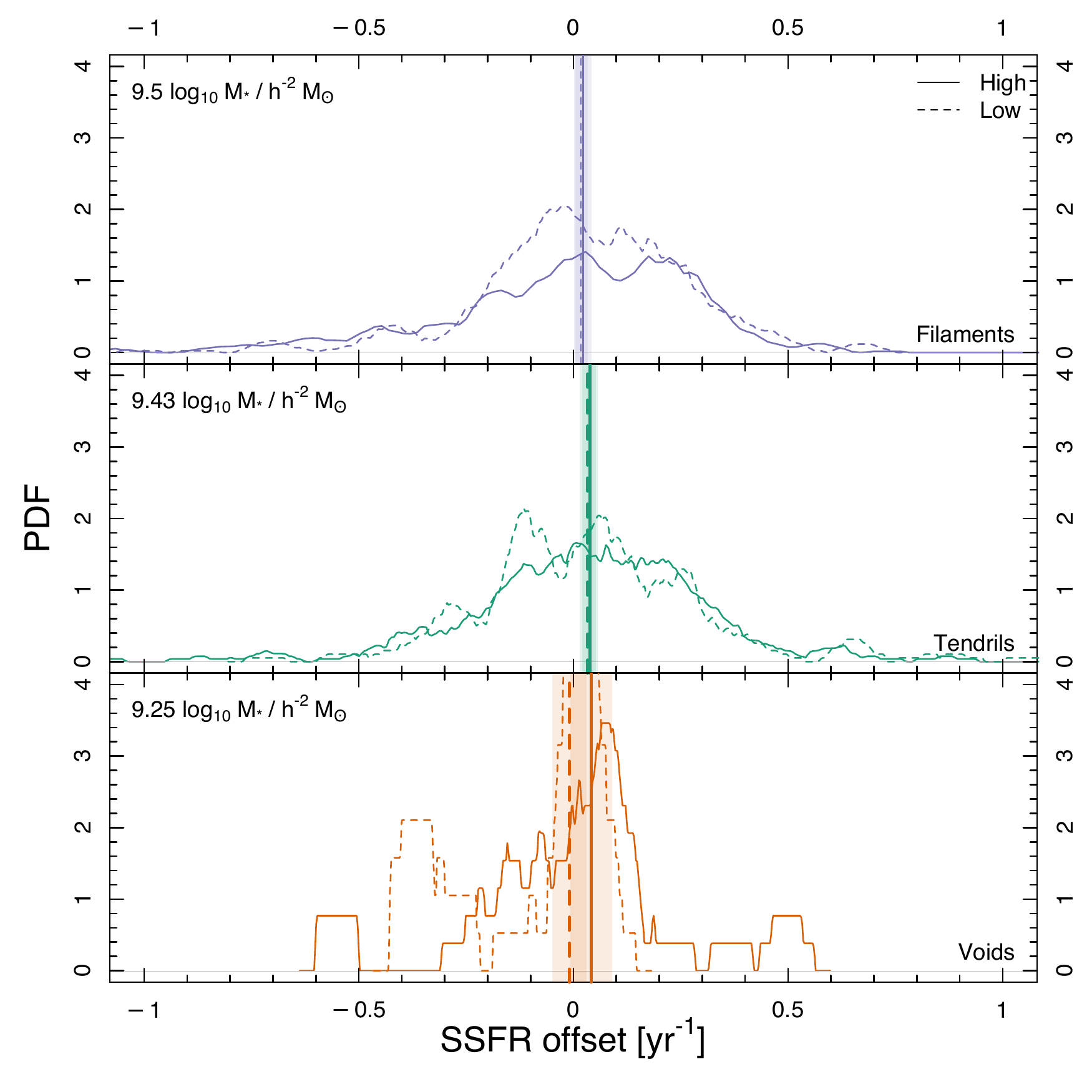}
	\caption{Specific star formation rates of spiral galaxies in different environments, subdivided into populations of low and high stellar mass (shown as dashed and solid lines respectively), with PDFs calculated by convolving the data with a rectangular kernel of bandwidth 0.1 dex. The vertical lines represent the median SSFR value for that population, and the shaded area shows the standard error on the median. While there is a small tendency for the SSFR of low mass galaxies to be higher in voids, the effect of stellar mass is greater than that of large scale environment. The median mass for each population, which divides the high and low mass sub-samples, is shown in the top left.}
	\label{fig:SSFRPDFbyEnv}
\end{figure}

We complement our initial sample of 1,799 unpaired and ungrouped spiral galaxies in filaments with an additional 946 and 67 in tendrils and voids respectively; in both cases the spiral galaxies are also star forming, ungrouped, and unpaired (hereafter these galaxies are referred to as the combined LSS sample). The calculation of the SSFR offset for the tendril and void populations is done in the same way as for the filaments, as described in Section \ref{sec:filoffset}. We further subdivide these galaxies by their stellar mass into two bins, above and below their respective population medians ($9.5,\; 9.43$, and $9.25 \; \log M_* / h^{-2} M_{\odot}$ for filaments, tendrils, and voids respectively). We show these distributions in \ref{fig:SSFRPDFbyEnv}, where in each panel SSFR offsets are shown by environment for galaxies above and below their populations' median stellar mass, with the median of each distribution and its standard error represented by the vertical line and shaded region in the corresponding colour. All three populations display very similar trends in SSFR offset; this result indicates that large-scale structure does not greatly impact the overall star formation rates of spiral galaxies.

\begin{figure*}
	\centering
	\includegraphics[width=\textwidth]{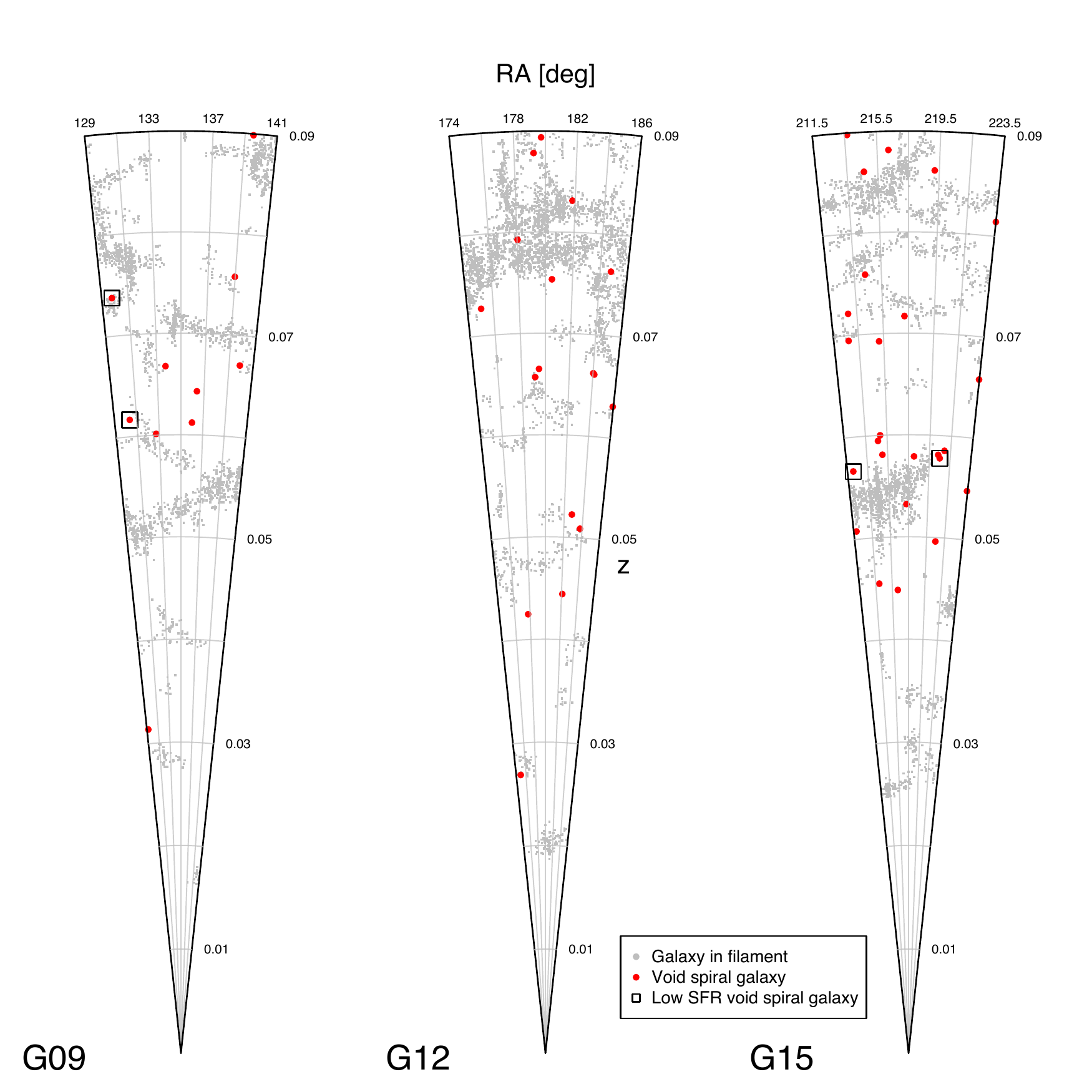}
	\caption{As with Figure \ref{fig:cones}, but displaying the positions of the spiral galaxies in voids for our sample as red circles. Note that any void galaxy within 1 degree of the survey edge, as well as with $z >= 0.085$ is not included in our analysis, as the large-scale structure classification of these galaxies as voids may not necessarily be acccurate. Void spiral galaxies that constitute the low star forming sub-population are shown in black squares.}
	\label{fig:voidcones}
\end{figure*}

Here we highlight the peculiar bimodality in SSFR offset seen for low mass void galaxies (as seen in the bottom panel of Fig. \ref{fig:SSFRPDFbyEnv}); this appears to be a sub-population of low mass galaxies that have strikingly low star formation rates. A jackknife resampling of the void galaxy sample confirms that this distribution is statistically significant. Visual inspection of these galaxies does not present an explanation for their lower star formation rates; nor are they physically clustered together. We display these galaxies in Figure \ref{fig:voidcones}, where we present the three equatorial GAMA fields as in Figure \ref{fig:cones}, this time highlighting our sample of void spiral galaxies in red. The sub-population of low mass star forming void galaxies are shown in black boxes.

\begin{figure}
	\centering
	\includegraphics[width=0.5\textwidth]{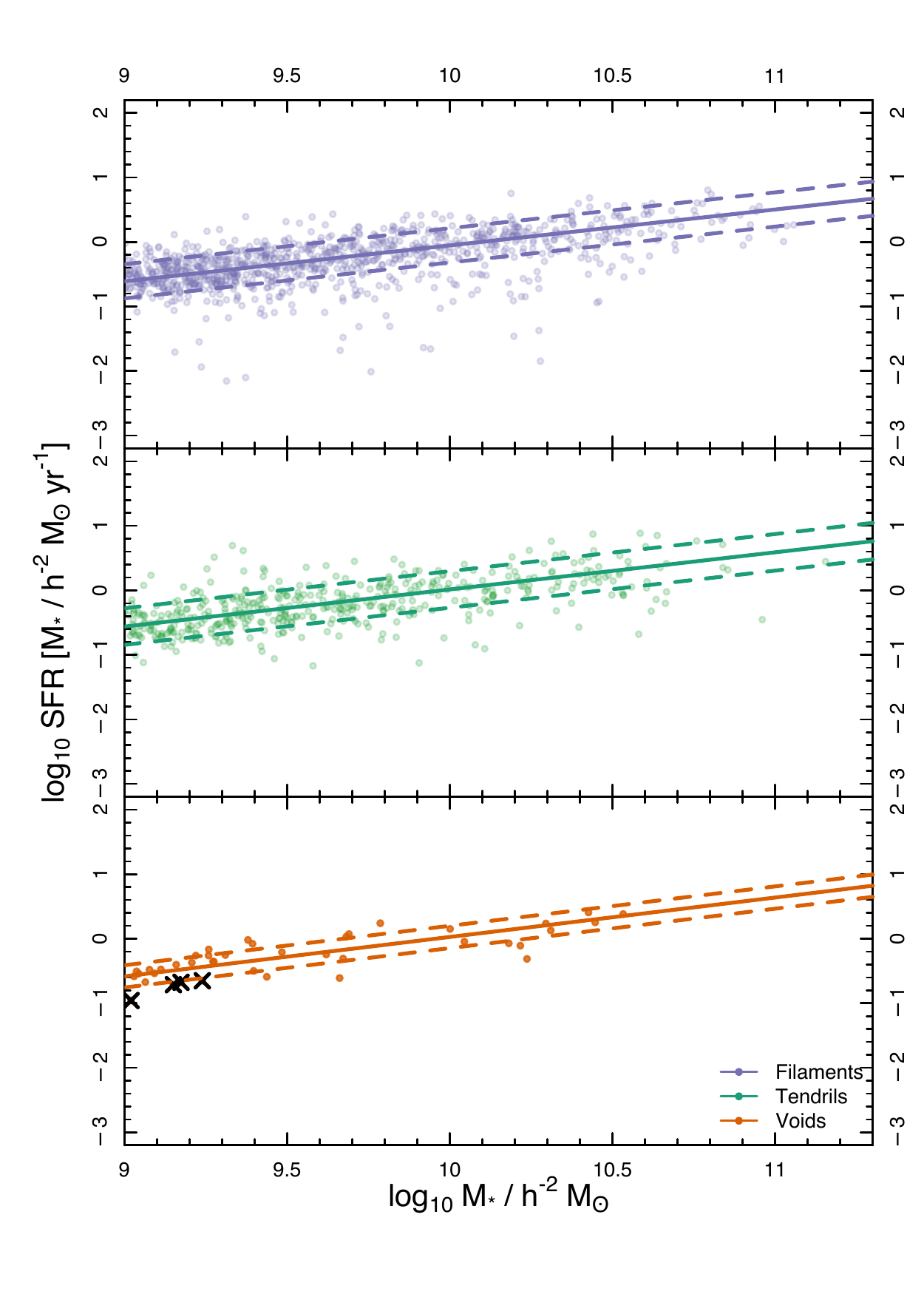}
	\caption{Star formation rates as a function of stellar mass for all galaxies in the combined LSS sample split by filaments, tendrils, and voids (top, middle, and bottom panels respectively). Model fits to the data are shown for each population as solid lines in their respective colours, with dashed lines representing the $1\sigma$ contour to the model. Black crosses in the bottom panel represent the galaxies that form the low mass, high SSFR offset peak seen in Figure \ref{fig:SSFRPDFbyEnv}. All three populations exhibit a similar SFR-M$_*$ relationship.}
	\label{fig:massvSFR}
\end{figure}

\begin{figure}
	\centering
	\includegraphics[width=0.5\textwidth]{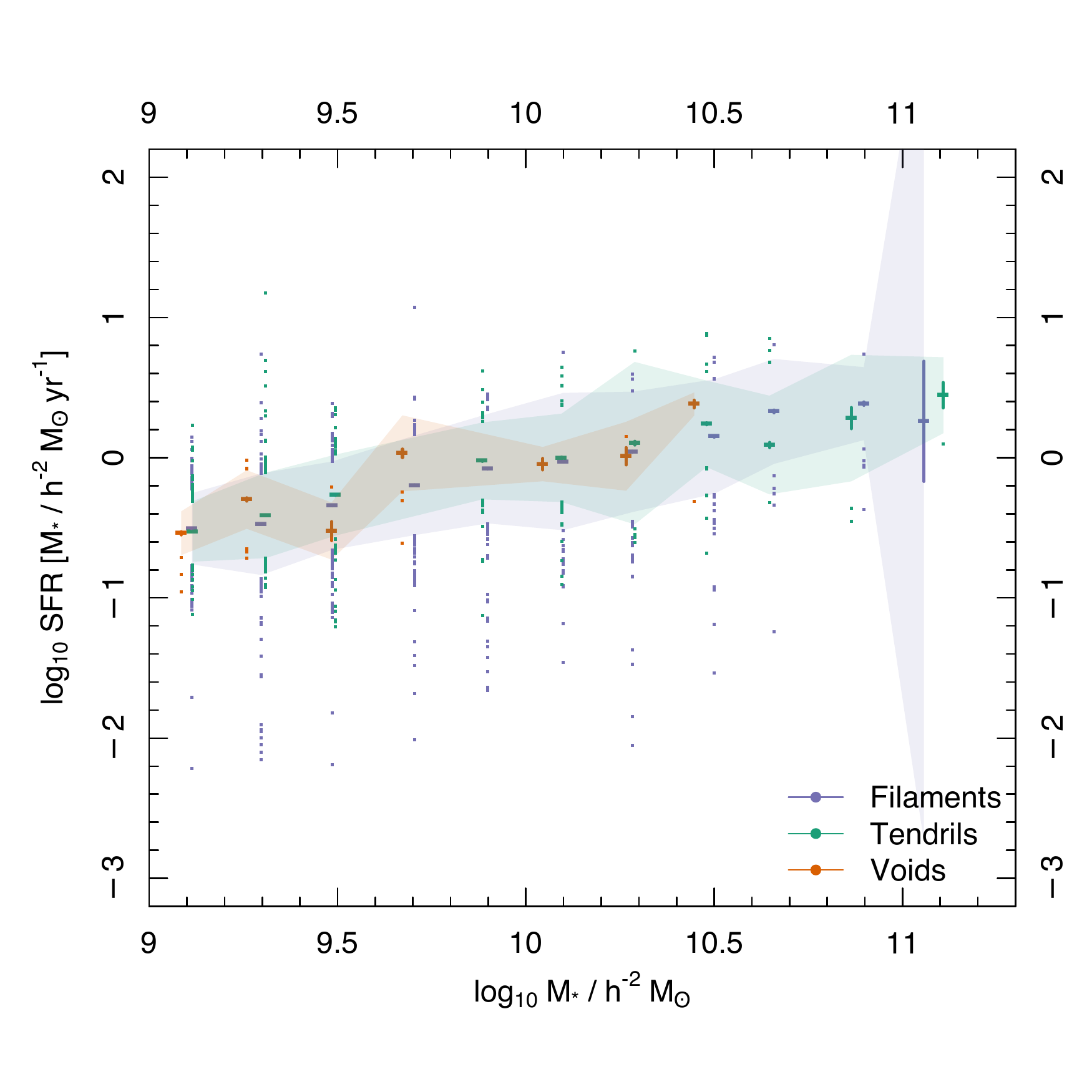}
	\caption{As with Fig. \ref{fig:massvSFR}, but with data binned into stellar mass bins of 0.2 dex. For each bin we show the median and its standard error as the solid lines; the variance as the shaded region, and outlier data points beyond the variance as individual points.}
	\label{fig:leftovers}
\end{figure}

\begin{figure*}
	\centering
	\includegraphics[width=\textwidth]{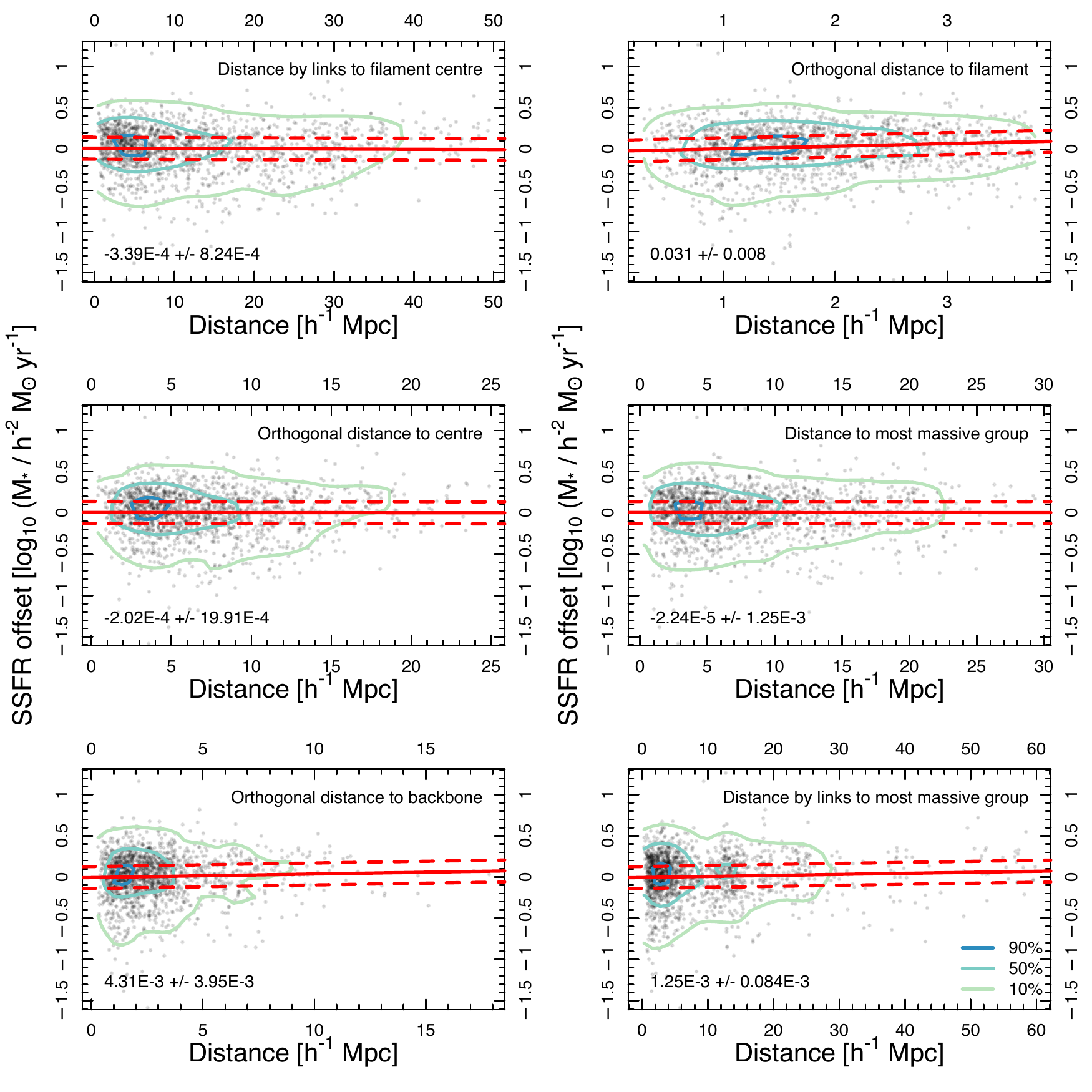}
	\caption{Offset in UV-derived specific star formation rates of spiral galaxies in filaments as a function of the 6 distance metrics described in Fig. \ref{fig:schematic}, with data and fits shown as in Fig. \ref{fig:dist_SFR_distToFil}. Aside from the orthogonal distance to the filament (top right), the gradients of the fits are statistically indistinguishable from being zero. The top right panel is the only trend with a non-zero slope at a $3\sigma$ significance and indicates that the star formation rates of spiral galaxies in filaments are higher closer to the edge of the filament when compared to the core of the filament. The detection of such an effect in ungrouped and unpaired spiral galaxies means that the most likely explanation for this change in star formation rates is a change in the availability of gas to these galaxies as a function of their orthogonal distance to the filament core.}
	\label{fig:dist_SFR_all}
\end{figure*}

The relative unimportance of large-scale structure in the star formation rates of spiral galaxies is also shown in Fig. \ref{fig:massvSFR}, where we plot SFR as a function of stellar mass for each galaxy in our combined LSS sample. As we show in Figs. \ref{fig:SSFRPDFbyEnv} and \ref{fig:massvSFR}, the median SSFR shows very little to no dependence on large-scale structure. However, the distributions indicate that environment may nevertheless play a secondary impact in terms of broadening the distribution, as seen in Figs. \ref{fig:SSFRPDFbyEnv} and \ref{fig:leftovers}. In the latter plot the variance in star formation rate as a function of stellar mass for galaxies in filaments, tendrils, and voids is shown. Galaxies in voids are shown to have a more narrow spread of star formation rates in each stellar mass bin. In Fig. \ref{fig:massvSFR} we also show, as black crosses, the population of void galaxies that generates the low mass, low SSFR offset in Figure \ref{fig:SSFRPDFbyEnv}. These low mass void spiral galaxies, while small in number, are forming far fewer stars than other void galaxies, and may represent a secondary mass-SFR main sequence for this population of galaxies. Environment does appear to play a role in the spread of star formation rates for a given mass bin, however.

\section{Discussion}

In this work we have sought out trends in star formation rate and stellar mass as a function of distance from a filament for a sample of ungrouped and unpaired spiral galaxies using data from the GAMA survey. Additionally, we have considered the instantaneous star formation of spiral galaxies in voids and tendrils contrasting these environments with the filament galaxies in our main sample. Furthermore, by making use of a mass complete morphologically selected sample as well as highly precise and accurate SFR measurements in the analysis presented in this work we have been able to quantify the subtle environmental dependencies in SFR at fixed stellar mass. This analysis has lead to a number of novel results which we will discuss in more detail below:

\begin{enumerate}
\item Spiral galaxies at the edges of filaments, on the borders of voids, have higher specific star formation rates (at fixed steller mass), and lower stellar masses compared to their counterparts at the cores of filaments. Specifically, our results indicate that spiral galaxies at the peripheries of filaments have greater access to reservoirs of star forming gas compared to their counterparts in the cores of filaments; and
\item Although the median relation between star formation rate and stellar mass is similar for filament, tendril and void spiral galaxies, the width of the distributions increases from voids to tendrils and filaments.
\end{enumerate}

\subsection{Star formation in filaments}

To investigate the impact the large-scale structure (in the form of filaments) has on (spiral) galaxies we considered the UV-derived SSFR of these galaxies as a function of their position relative to the surrounding large-scale structure, i.e. their host filament using 6 different distance metrics (Fig. \ref{fig:dist_SFR_distToFil}). Of these we find a statistically significant trend in SSFR with position for one: the orthogonal distance to the filament, which is shown in Fig. \ref{fig:dist_SFR_distToFil}. A corresponding trend is found for the stellar mass of the galaxies, with galaxies closer to the cylindrical core being found to be more massive on average (Fig. \ref{fig:dist_mass_distToFil}). Our results therefore indicate that spiral galaxies close to the cylindrical core of a filament have a very slight tendency to have lower specific star formation rates and higher stellar masses. When comparing the median SSFR - $M_*$ relations for spiral galaxies in filaments voids and tendrils in Fig. \ref{fig:massvSFR}, however, we find no significant difference. This result confirms the findings of \citet{Alpaslan2015} and Grootes et al. (in prep) that stellar mass plays a decisive role in determining galaxy properties, and that environmental effects such as those arising from the large-scale structure are second order perturbations on this dominant trend.

The observed changes in stellar populations for these spiral galaxies as a function of their position with respect to a filament may be caused by a number of different physical processes. At a fundamental level, however, in order to support star formation at the observed level for extended periods, spiral galaxies must accrete gas from their surrounding intergalactic medium (IGM) in order to replenish the interstellar medium consumed by star formation (as well as that lost to outflows from the galaxy).

For isolated galaxies at the centre of halos with $M_halo \lesssim 10^{12} M_{\odot}$ \citet{Dekel2006} have argued that this fueling predominantly takes the form of smooth cold flows of gas penetrating the virial radius of the dark matter (DM) halo to reach the galaxy, leading to the formation of a a disk and efficient star formation. Furthermore, recent work by \citet{Pichon2011,Codis2015,Welker2015} indicates that the preferential directions for these cold flows are determined by the large scale filamentary DM structure, with the filaments being fed by flows of gas from voids and walls, which is then advected onto the DM halo of the galaxy as a cold flow with high angular momentum, thus further spinning up the galaxy and supporting the disk.

The gas being accreted into the filament, however, is heated by the cylindrical shock surrounding the filament as well as by shocks within the filament. Furthermore, as recently demonstrated by \citet{2013MNRAS.429.3353N} cool dense flows into a more diffuse hot medium may be efficiently heated and dispersed. Thus, the rate at which gas can be accreted onto a (spiral) galaxy in a filament will depend on (i) the degree to which the cold flows entering the filament retain their coherence and remain connected to the halo of the galaxy and (ii) the cooling timescale of the shock-heated gas which depends sensitively on the temperature.

\begin{figure*}
	\centering
	\includegraphics[width=\textwidth]{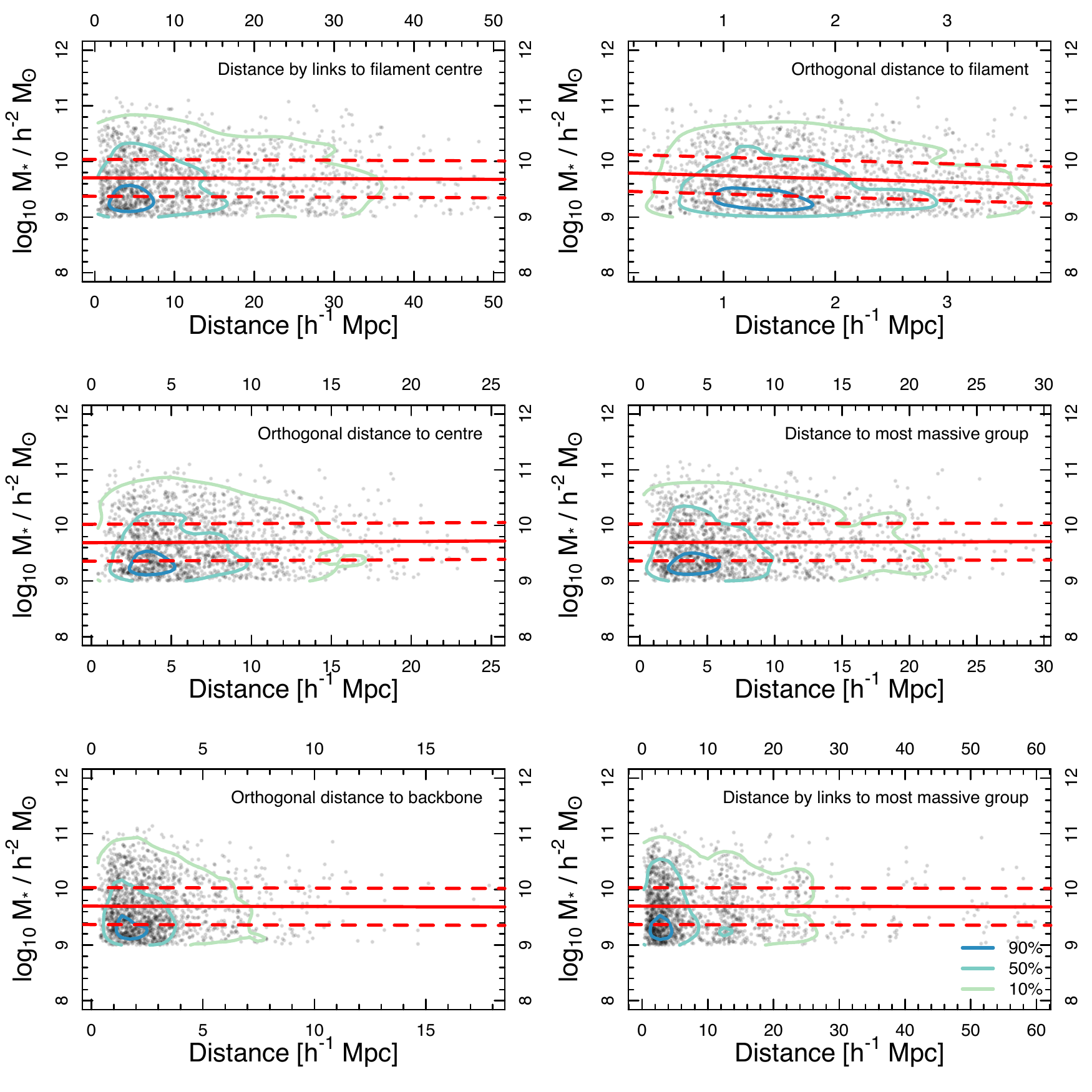}
	\caption{Stellar masses of spiral galaxies in filaments as a function of the 6 distance metrics described in Fig. \ref{fig:schematic}, with data and fits shown as Fig. \ref{fig:dist_mass_distToFil}. We detect a statistically significant downward trend in stellar mass as a function of orthogonal distance to the filament (topright), while all other fits are statistically consistent with having a gradient of zero.}
	\label{fig:dist_mass_all}
\end{figure*}

Numerical simulations of the IGM in and around filaments \citep{Dolag2006,2015arXiv151005388Z} find a denser but simultaneously hotter IGM in filaments, with both temperature and density increasing towards the centre, while at smaller scales the temperature of gas inside relaxed clusters decreases towards the cluster outskirts \citep{Kravtsov2012}. Furthermore recent work by \citet{Benitez-Llambay2013} suggests that low mass haloes can have their gas stripped by ram pressure caused by the cosmic web. In combination, these trends may limit the inflow of accretable gas, providing a natural explanation for our empirical findings. With ongoing star formation in galaxies, such a scenario would also provide a natural explanation for the trend in stellar mass seen in Fig. \ref{fig:dist_mass_all}. More detailed modeling and further observations will be necessary to investigate this scenario, though such work is beyond the scope of this paper.

Other possible explanations may be that the observed trends in SFR mirror the availability of gas to galaxies in filaments as a function of their position; these galaxies may simply be exposed to less gas with which to form stars, or are cut off from the resupply of gas and deplete their gas reservoirs as they migrate to the core of the filament (but see Grootes et al. (in prep) for evidence that this explanation is unlikely to be the case). Notably, in our analysis we have only found evidence for variations in the SFR and stellar mass perpendicular to the filament rather than along it. It seems likely that the gradient in the gravitational potential, and hence the accompanying changes in the thermodynamic properties or availability of the IGM is largest perpendicular to the filament. We must, however, also consider the possibility that our metrics struggle to capture the complex geometry of filamentary structures (evident in Figs. \ref{fig:cones} and \ref{fig:schematic}), which will more severely impact metrics designed to trace this geometry over large distances (particularly those designed to follow links in the minimal spanning tree), causing a potential loss in signal. We are, further, only probing the properties of the IGM/flow of gas by tracking star formation rates in spiral galaxies; ultimately, a detailed study of mass transport along the cosmic web will be rendered possible through the advent of next generation radio surveys to be conducted at the ASKAP, SKA or the MeerKAT.

\subsection{Star formation in voids}

As previously stated, and shown in detail in Figs. \ref{fig:massvSFR} and \ref{fig:leftovers}, the average SFR - $M_*$ relations for spiral galaxies in filaments, tendrils and voids are highly similar. However, as shown in Fig. \ref{fig:SSFRPDFbyEnv} the distribution of SSFR around the relation at fixed stellar mass is much more peaked in voids than in filaments and tendrils. This difference may be taken to be indicative of the the star formation histories in the former environment being much more uniform than in the other denser environments, where processes such as galaxy-galaxy interactions and different interactions of the galaxies with the changing surrounding IGM may lead to a diversification of the star formation history in comparison to the situation of void spiral galaxies, where star formation may be expected to be largely determined by the accretion of ambient IGM onto the host dark matter halo of the galaxy alone.

Finally, we note that for spiral galaxies in the mass range $10^9 - 10^{9.5}$ $M_{\odot}$ in voids, one finds a clear bimodality in the distribution of SSFR with a small, but statistically significant secondary peak in the distribution offset to lower sFSR by 0.5 dex. Considering the positions of these sources on the SFR - $M_*$ relation shown in Fig. \ref{fig:massvSFR}, we find that this peak appears to trace a sequence parallel to that of the bulk of the void spiral galaxies, but offset towards lower star formation. Although the sample size is small, the observed bimodality remains robustly visible under jackknife sampling with a variation of sizes of the jackknifed sample. We therefore conclude the result to be robust in nature and physical. While the sample is too small in size for a more in depth analysis, we limit ourselves to drawing attention to the existence of this unexpected population of void spiral galaxies. We speculate that this feature may arise due to a more stochastic nature of star formation in the lowest mass galaxies, with the more diverse star formation history in higher density environments masking this effect in tendrils and filaments in the relevant stellar mass range.

\section{Summary}

As expected, the stellar mass of a spiral galaxy plays a dominant role in determining its star formation rate. However, by making use of a sensitive analysis technique we have been able to identify environmental trends superimposed on this relation between star formation and stellar mass. In line with expectations based on the picture of large scale flows of gas onto filaments we find the specific star formation rates of spiral galaxies in filaments to be higher at fixed stellar mass on the periphery of the filament compared to its core, and also find the distribution of stellar masses of filament spiral galaxies to be skewed towards higher mass systems at the core. Extending our analysis to tendrils and voids, we have identified a significant bimodality in the distribution of SSFR of low mass void spiral galaxies, which we speculate may be driven by the stochastic nature of star formation in these objects.

The results presented in this work must be considered as promising evidence for the complex interplay between galaxies and the cosmic web, as well as the influence of large-scale structure on galaxy evolution. They present a clear point of departure for future studies with new facitilies and surveys capable of probing the relevant processes more directly and in greater detail.

\section*{Acknowledgements}

MA is funded by an appointment to the NASA Postdoctoral Program at Ames Research Centre, administered by Oak Ridge Associated Universities through a contract with NASA. MALL acknowledges support from UNAM through the PAPIIT project IA101315. The authors would like to thank the anonymous referee, whose contributions have helped to improve this work.

GAMA is a joint European-Australasian project based around a spectroscopic campaign using the Anglo-Australian Telescope. The GAMA input catalogue is based on data taken from the Sloan Digital Sky Survey and the UKIRT Infrared Deep Sky Survey. Complementary imaging of the GAMA regions is being obtained by a number of independent survey programs including GALEX MIS, VST KiDS, VISTA VIKING, WISE, Herschel-ATLAS, GMRT and ASKAP providing UV to radio coverage. GAMA is funded by the STFC (UK), the ARC (Australia), the AAO, and the participating institutions. The GAMA website is http://www.gama-survey.org/. The VISTA VIKING data used in this paper is based on observations made with ESO Telescopes at the La Silla Paranal Observatory under programme ID 179.A-2004.

\footnotesize
\bibliographystyle{mnras}
\setlength{\bibhang}{2.0em}
\setlength{\labelwidth}{0.0em}
\bibliography{alongFilament.bib}

\begin{thebibliography}{}
\makeatletter
\relax
\def\mn@urlcharsother{\let\do\@makeother \do\$\do\&\do\#\do\^\do\_\do\%\do\~}
\def\mn@doi{\begingroup\mn@urlcharsother \@ifnextchar [ {\mn@doi@}
  {\mn@doi@[]}}
\def\mn@doi@[#1]#2{\def\@tempa{#1}\ifx\@tempa\@empty \href
  {http://dx.doi.org/#2} {doi:#2}\else \href {http://dx.doi.org/#2} {#1}\fi
  \endgroup}
\def\mn@eprint#1#2{\mn@eprint@#1:#2::\@nil}
\def\mn@eprint@arXiv#1{\href {http://arxiv.org/abs/#1} {{\tt arXiv:#1}}}
\def\mn@eprint@dblp#1{\href {http://dblp.uni-trier.de/rec/bibtex/#1.xml}
  {dblp:#1}}
\def\mn@eprint@#1:#2:#3:#4\@nil{\def\@tempa {#1}\def\@tempb {#2}\def\@tempc
  {#3}\ifx \@tempc \@empty \let \@tempc \@tempb \let \@tempb \@tempa \fi \ifx
  \@tempb \@empty \def\@tempb {arXiv}\fi \@ifundefined
  {mn@eprint@\@tempb}{\@tempb:\@tempc}{\expandafter \expandafter \csname
  mn@eprint@\@tempb\endcsname \expandafter{\@tempc}}}

\bibitem[\protect\citeauthoryear{Alpaslan et~al.,}{Alpaslan
  et~al.}{2012}]{Alpaslan2012}
Alpaslan M.,  et~al., 2012, \mn@doi [Monthly Notices of the Royal Astronomical
  Society] {10.1111/j.1365-2966.2012.21020.x}, 426, 2832

\bibitem[\protect\citeauthoryear{Alpaslan et~al.,}{Alpaslan
  et~al.}{2014a}]{Alpaslan2013a}
Alpaslan M.,  et~al., 2014a, \mn@doi [Monthly Notices of the Royal Astronomical
  Society] {10.1093/mnras/stt2136}, 438, 177

\bibitem[\protect\citeauthoryear{Alpaslan et~al.,}{Alpaslan
  et~al.}{2014b}]{Alpaslan2014b}
Alpaslan M.,  et~al., 2014b, \mn@doi [Monthly Notices of the Royal Astronomical
  Society: Letters] {10.1093/mnrasl/slu019}, 440, L106

\bibitem[\protect\citeauthoryear{Alpaslan et~al.,}{Alpaslan
  et~al.}{2015}]{Alpaslan2015}
Alpaslan M.,  et~al., 2015, \mn@doi [Monthly Notices of the Royal Astronomical
  Society] {10.1093/mnras/stv1176}, 451, 3249

\bibitem[\protect\citeauthoryear{Arag{\'{o}}n-Calvo, Jones, van~de Weygaert  \&
  van~der Hulst}{Arag{\'{o}}n-Calvo et~al.}{2007}]{Aragon-Calvo2007}
Arag{\'{o}}n-Calvo M.~A.,  Jones B. J.~T.,  van~de Weygaert R.,   van~der Hulst
  J.~M.,  2007, Astronomy and Astrophysics, 474, 315

\bibitem[\protect\citeauthoryear{Arag\'{o}n-Calvo, van~de Weygaert  \&
  Jones}{Arag\'{o}n-Calvo et~al.}{2010}]{Aragon-Calvo2010}
Arag\'{o}n-Calvo M.~A.,  van~de Weygaert R.,   Jones B. J.~T.,  2010, Monthly
  Notices of the Royal Astronomical Society, 408, 2163

\bibitem[\protect\citeauthoryear{Ben\'{\i}tez-Llambay, Navarro, Abadi,
  Gottl\"{o}ber, Yepes, Hoffman  \& Steinmetz}{Ben\'{\i}tez-Llambay
  et~al.}{2013}]{Benitez-Llambay2013}
Ben\'{\i}tez-Llambay A.,  Navarro J.~F.,  Abadi M.~G.,  Gottl\"{o}ber S.,
  Yepes G.,  Hoffman Y.,   Steinmetz M.,  2013, \mn@doi [The Astrophysical
  Journal] {10.1088/2041-8205/763/2/L41}, 763, L41

\bibitem[\protect\citeauthoryear{{Blanton} \& {Roweis}}{{Blanton} \&
  {Roweis}}{2007}]{BLANTON2007}
{Blanton} M.~R.,  {Roweis} S.,  2007, \mn@doi [\aj] {10.1086/510127}, \href
  {http://adsabs.harvard.edu/abs/2007AJ....133..734B} {133, 734}

\bibitem[\protect\citeauthoryear{Bond, Kofman  \& Pogosyan}{Bond
  et~al.}{1996}]{Bond1996}
Bond J.~R.,  Kofman L.,   Pogosyan D.,  1996, Nature, 380, 603

\bibitem[\protect\citeauthoryear{Brough, Forbes, Kilborn, Couch  \&
  Colless}{Brough et~al.}{2006}]{Brough2006}
Brough S.,  Forbes D.~A.,  Kilborn V.~A.,  Couch W.,   Colless M.,  2006,
  \mn@doi [Monthly Notices of the Royal Astronomical Society]
  {10.1111/j.1365-2966.2006.10387.x}, 369, 1351

\bibitem[\protect\citeauthoryear{Brough et~al.,}{Brough
  et~al.}{2013}]{Brough2013}
Brough S.,  et~al., 2013, \mn@doi [Monthly Notices of the Royal Astronomical
  Society] {10.1093/mnras/stt1489}, 435, 2903

\bibitem[\protect\citeauthoryear{Brown et~al.,}{Brown et~al.}{2008}]{Brown2008}
Brown M. J.~I.,  et~al., 2008, \mn@doi [The Astrophysical Journal]
  {10.1086/589538}, 682, 937

\bibitem[\protect\citeauthoryear{Cautun, van~de Weygaert  \& Jones}{Cautun
  et~al.}{2012}]{Cautun2012}
Cautun M.,  van~de Weygaert R.,   Jones B. J.~T.,  2012, \mn@doi [Monthly
  Notices of the Royal Astronomical Society] {10.1093/mnras/sts416}, 429, 1286

\bibitem[\protect\citeauthoryear{Cautun, van~de Weygaert, Jones  \&
  Frenk}{Cautun et~al.}{2014}]{Cautun2014}
Cautun M.,  van~de Weygaert R.,  Jones B. J.~T.,   Frenk C.~S.,  2014, \mn@doi
  [Monthly Notices of the Royal Astronomical Society] {10.1093/mnras/stu768},
  441, 2923

\bibitem[\protect\citeauthoryear{{Chabrier}}{{Chabrier}}{2003}]{CHABRIER2003}
{Chabrier} G.,  2003, \mn@doi [\pasp] {10.1086/376392}, \href
  {http://adsabs.harvard.edu/abs/2003PASP..115..763C} {115, 763}

\bibitem[\protect\citeauthoryear{Chen et~al.,}{Chen et~al.}{2015}]{Chen2015}
Chen Y.-C.,  et~al., 2015, p.~12

\bibitem[\protect\citeauthoryear{Codis, Pichon  \& Pogosyan}{Codis
  et~al.}{2015}]{Codis2015}
Codis S.,  Pichon C.,   Pogosyan D.,  2015, \mn@doi [Monthly Notices of the
  Royal Astronomical Society] {10.1093/mnras/stv1570}, 452, 3369

\bibitem[\protect\citeauthoryear{Colberg}{Colberg}{2007}]{Colberg2007}
Colberg J.~M.,  2007, \mn@doi [Monthly Notices of the Royal Astronomical
  Society] {10.1111/j.1365-2966.2006.11312.x}, 375, 337

\bibitem[\protect\citeauthoryear{Darvish, Sobral, Mobasher, Scoville, Best,
  Sales  \& Smail}{Darvish et~al.}{2014}]{Darvish2014}
Darvish B.,  Sobral D.,  Mobasher B.,  Scoville N.~Z.,  Best P.,  Sales L.~V.,
   Smail I.,  2014, \mn@doi [The Astrophysical Journal]
  {10.1088/0004-637X/796/1/51}, 796, 51

\bibitem[\protect\citeauthoryear{Davies et~al.,}{Davies
  et~al.}{2015}]{Davies2015}
Davies L. J.~M.,  et~al., 2015, \mn@doi [Monthly Notices of the Royal
  Astronomical Society] {10.1093/mnras/stv1241}, 452, 616

\bibitem[\protect\citeauthoryear{Dekel \& Birnboim}{Dekel \&
  Birnboim}{2006}]{Dekel2006}
Dekel A.,  Birnboim Y.,  2006, \mn@doi [Monthly Notices of the Royal
  Astronomical Society] {10.1111/j.1365-2966.2006.10145.x}, 368, 2

\bibitem[\protect\citeauthoryear{Dolag, Meneghetti, Moscardini, Rasia  \&
  Bonaldi}{Dolag et~al.}{2006}]{Dolag2006}
Dolag K.,  Meneghetti M.,  Moscardini L.,  Rasia E.,   Bonaldi A.,  2006,
  \mn@doi [Monthly Notices of the Royal Astronomical Society]
  {10.1111/j.1365-2966.2006.10511.x}, 370, 656

\bibitem[\protect\citeauthoryear{Doroshkevich, Tucker, Allam  \&
  Way}{Doroshkevich et~al.}{2004}]{Doroshkevich2004}
Doroshkevich A.,  Tucker D.~L.,  Allam S.,   Way M.~J.,  2004, \mn@doi
  [Astronomy and Astrophysics] {10.1051/0004-6361:20031780}, 418, 7

\bibitem[\protect\citeauthoryear{Driver et~al.,}{Driver
  et~al.}{2009}]{Driver2009}
Driver S.~P.,  et~al., 2009, Astronomy and Geophysics, 50, 12

\bibitem[\protect\citeauthoryear{Driver et~al.,}{Driver
  et~al.}{2011}]{Driver2011}
Driver S.~P.,  et~al., 2011, Monthly Notices of the Royal Astronomical Society,
  413, 971

\bibitem[\protect\citeauthoryear{{Driver} et~al.,}{{Driver}
  et~al.}{2016}]{Driver2016}
{Driver} S.~P.,  et~al., 2016, \mn@doi [\mnras] {10.1093/mnras/stv2505}, \href
  {http://adsabs.harvard.edu/abs/2016MNRAS.455.3911D} {455, 3911}

\bibitem[\protect\citeauthoryear{Eardley et~al.,}{Eardley
  et~al.}{2015}]{Eardley2015}
Eardley E.,  et~al., 2015, \mn@doi [Monthly Notices of the Royal Astronomical
  Society] {10.1093/mnras/stv237}, 448, 3665

\bibitem[\protect\citeauthoryear{El-Ad \& Piran}{El-Ad \&
  Piran}{1997}]{El-Ad1997}
El-Ad H.,  Piran T.,  1997, The Astrophysical Journal, 491, 421

\bibitem[\protect\citeauthoryear{Ellison, Patton, Simard  \&
  McConnachie}{Ellison et~al.}{2008}]{Ellison2008}
Ellison S.~L.,  Patton D.~R.,  Simard L.,   McConnachie A.~W.,  2008, \mn@doi
  [The Astronomical Journal] {10.1088/0004-6256/135/5/1877}, 135, 1877

\bibitem[\protect\citeauthoryear{Fadda, Biviano, Marleau, Storrie-Lombardi  \&
  Durret}{Fadda et~al.}{2008}]{Fadda2008}
Fadda D.,  Biviano A.,  Marleau F.~R.,  Storrie-Lombardi L.~J.,   Durret F.,
  2008, The Astrophysical Journal Letters, 672, L9

\bibitem[\protect\citeauthoryear{Forero-Romero, Hoffman, Gottl\"{o}ber, Klypin
  \& Yepes}{Forero-Romero et~al.}{2009}]{Forero-Romero2009}
Forero-Romero J.~E.,  Hoffman Y.,  Gottl\"{o}ber S.,  Klypin A.,   Yepes G.,
  2009, Monthly Notices of the Royal Astronomical Society, 396, 1815

\bibitem[\protect\citeauthoryear{Gilbank, Baldry, Balogh, Glazebrook  \&
  Bower}{Gilbank et~al.}{2010}]{Gilbank2010}
Gilbank D.~G.,  Baldry I.~K.,  Balogh M.~L.,  Glazebrook K.,   Bower R.~G.,
  2010, \mn@doi [Monthly Notices of the Royal Astronomical Society]
  {10.1111/j.1365-2966.2010.16640.x}, 405, no

\bibitem[\protect\citeauthoryear{Gray \& Scannapieco}{Gray \&
  Scannapieco}{2013}]{Gray2013}
Gray W.~J.,  Scannapieco E.,  2013, \mn@doi [The Astrophysical Journal]
  {10.1088/0004-637X/768/2/174}, 768, 174

\bibitem[\protect\citeauthoryear{Grootes, Tuffs, Popescu, Robotham, Seibert  \&
  Kelvin}{Grootes et~al.}{2013}]{Grootes2013}
Grootes M.~W.,  Tuffs R.~J.,  Popescu C.~C.,  Robotham A. S.~G.,  Seibert M.,
  Kelvin L.~S.,  2013, \mn@doi [Monthly Notices of the Royal Astronomical
  Society] {10.1093/mnras/stt2184}, 437, 3883

\bibitem[\protect\citeauthoryear{{Grootes}, {Tuffs}, {Popescu}, {Robotham},
  {Seibert}  \& {Kelvin}}{{Grootes} et~al.}{2014}]{GROOTES2014}
{Grootes} M.~W.,  {Tuffs} R.~J.,  {Popescu} C.~C.,  {Robotham} A.~S.~G.,
  {Seibert} M.,   {Kelvin} L.~S.,  2014, \mn@doi [\mnras]
  {10.1093/mnras/stt2184}, \href
  {http://adsabs.harvard.edu/abs/2014MNRAS.437.3883G} {437, 3883}

\bibitem[\protect\citeauthoryear{Gunawardhana et~al.,}{Gunawardhana
  et~al.}{2013}]{Gunawardhana2013}
Gunawardhana M. L.~P.,  et~al., 2013, \mn@doi [Monthly Notices of the Royal
  Astronomical Society] {10.1093/mnras/stt890}, 433, 2764

\bibitem[\protect\citeauthoryear{Hahn, Porciani, Carollo  \& Dekel}{Hahn
  et~al.}{2007}]{Hahn2007}
Hahn O.,  Porciani C.,  Carollo C.~M.,   Dekel A.,  2007, Monthly Notices of
  the Royal Astronomical Society, 375, 489

\bibitem[\protect\citeauthoryear{Hopkins et~al.,}{Hopkins
  et~al.}{2013}]{Hopkins2013}
Hopkins A.~M.,  et~al., 2013, \mn@doi [Monthly Notices of the Royal
  Astronomical Society] {10.1093/mnras/stt030}, 430, 2047

\bibitem[\protect\citeauthoryear{Kauffmann et~al.,}{Kauffmann
  et~al.}{2003}]{Kauffmann2003}
Kauffmann G.,  et~al., 2003, \mn@doi [Monthly Notices of the Royal Astronomical
  Society] {10.1111/j.1365-2966.2003.07154.x}, 346, 1055

\bibitem[\protect\citeauthoryear{Kelvin et~al.,}{Kelvin
  et~al.}{2012}]{Kelvin2012}
Kelvin L.~S.,  et~al., 2012, \mn@doi [Monthly Notices of the Royal Astronomical
  Society] {10.1111/j.1365-2966.2012.20355.x}, 421, 1007

\bibitem[\protect\citeauthoryear{{Kennicutt}}{{Kennicutt}}{1998}]{KENNICUTT1998}
{Kennicutt} Jr. R.~C.,  1998, \mn@doi [\araa] {10.1146/annurev.astro.36.1.189},
  \href {http://adsabs.harvard.edu/abs/1998ARA%26A..36..189K} {36, 189}

\bibitem[\protect\citeauthoryear{Kere{\v s}, Katz, Weinberg  \& Dave}{Kere{\v
  s} et~al.}{2005}]{Kere2005}
Kere{\v s} D.,  Katz N.,  Weinberg D.~H.,   Dave R.,  2005, \mn@doi [Monthly
  Notices of the Royal Astronomical Society]
  {10.1111/j.1365-2966.2005.09451.x}, 363, 2

\bibitem[\protect\citeauthoryear{Kravtsov \& Borgani}{Kravtsov \&
  Borgani}{2012}]{Kravtsov2012}
Kravtsov A.~V.,  Borgani S.,  2012, \mn@doi [Annual Review of Astronomy and
  Astrophysics] {10.1146/annurev-astro-081811-125502}, 50, 353

\bibitem[\protect\citeauthoryear{Kreckel, {Ryan Joung}  \& Cen}{Kreckel
  et~al.}{2011}]{Kreckel2011}
Kreckel K.,  {Ryan Joung} M.,   Cen R.,  2011, \mn@doi [The Astrophysical
  Journal] {10.1088/0004-637X/735/2/132}, 735, 132

\bibitem[\protect\citeauthoryear{Kreckel, Platen, Arag{\'{o}}n-Calvo, van
  Gorkom, van~de Weygaert, van~der Hulst  \& Beygu}{Kreckel
  et~al.}{2012}]{Kreckel2012}
Kreckel K.,  Platen E.,  Arag{\'{o}}n-Calvo M.~A.,  van Gorkom J.~H.,  van~de
  Weygaert R.,  van~der Hulst J.~M.,   Beygu B.,  2012, \mn@doi [The
  Astronomical Journal] {10.1088/0004-6256/144/1/16}, 144, 16

\bibitem[\protect\citeauthoryear{Lintott et~al.,}{Lintott
  et~al.}{2011}]{Lintott2011}
Lintott C.,  et~al., 2011, \mn@doi [Monthly Notices of the Royal Astronomical
  Society] {10.1111/j.1365-2966.2010.17432.x}, 410, 166

\bibitem[\protect\citeauthoryear{Liske et~al.,}{Liske et~al.}{2015}]{Liske2015}
Liske Â.,  et~al., 2015, eprint arXiv:1506.08222

\bibitem[\protect\citeauthoryear{Mahajan, Raychaudhury  \& Pimbblet}{Mahajan
  et~al.}{2012}]{Mahajan2012}
Mahajan S.,  Raychaudhury S.,   Pimbblet K.~A.,  2012, \mn@doi [Monthly Notices
  of the Royal Astronomical Society] {10.1111/j.1365-2966.2012.22059.x}, 427,
  1252

\bibitem[\protect\citeauthoryear{{Martin} et~al.,}{{Martin}
  et~al.}{2005}]{MARTIN2005}
{Martin} D.~C.,  et~al., 2005, \mn@doi [\apjl] {10.1086/426387}, \href
  {http://adsabs.harvard.edu/abs/2005ApJ...619L...1M} {619, L1}

\bibitem[\protect\citeauthoryear{{Morrissey} et~al.,}{{Morrissey}
  et~al.}{2007}]{MORRISSEY2007}
{Morrissey} P.,  et~al., 2007, \mn@doi [\apjs] {10.1086/520512}, \href
  {http://adsabs.harvard.edu/abs/2007ApJS..173..682M} {173, 682}

\bibitem[\protect\citeauthoryear{Nelder \& Mead}{Nelder \&
  Mead}{1965}]{Nelder1965}
Nelder J.~A.,  Mead R.,  1965, \mn@doi [The Computer Journal]
  {10.1093/comjnl/7.4.308}, 7, 308

\bibitem[\protect\citeauthoryear{{Nelson}, {Vogelsberger}, {Genel}, {Sijacki},
  {Kere{\v s}}, {Springel}  \& {Hernquist}}{{Nelson}
  et~al.}{2013}]{2013MNRAS.429.3353N}
{Nelson} D.,  {Vogelsberger} M.,  {Genel} S.,  {Sijacki} D.,  {Kere{\v s}} D.,
  {Springel} V.,   {Hernquist} L.,  2013, \mn@doi [\mnras]
  {10.1093/mnras/sts595}, \href {http://esoads.eso.org/abs/2013MNRAS.429.3353N}
  {429, 3353}

\bibitem[\protect\citeauthoryear{Neyrinck}{Neyrinck}{2008}]{Neyrinck2008}
Neyrinck M.~C.,  2008, \mn@doi [Monthly Notices of the Royal Astronomical
  Society] {10.1111/j.1365-2966.2008.13180.x}, 386, 2101

\bibitem[\protect\citeauthoryear{{Noeske} et~al.,}{{Noeske}
  et~al.}{2007}]{NOESKE2007}
{Noeske} K.~G.,  et~al., 2007, \mn@doi [\apjl] {10.1086/517926}, \href
  {http://adsabs.harvard.edu/abs/2007ApJ...660L..43N} {660, L43}

\bibitem[\protect\citeauthoryear{Penny et~al.,}{Penny et~al.}{2015}]{Penny2015}
Penny S.~J.,  et~al., 2015, \mn@doi [Monthly Notices of the Royal Astronomical
  Society] {10.1093/mnras/stv1926}, 453, 3520

\bibitem[\protect\citeauthoryear{Pichon, Pogosyan, Kimm, Slyz, Devriendt  \&
  Dubois}{Pichon et~al.}{2011}]{Pichon2011}
Pichon C.,  Pogosyan D.,  Kimm T.,  Slyz A.,  Devriendt J.,   Dubois Y.,  2011,
  \mn@doi [Monthly Notices of the Royal Astronomical Society]
  {10.1111/j.1365-2966.2011.19640.x}, 418, 2493

\bibitem[\protect\citeauthoryear{Platen, {Van De Weygaert}  \& Jones}{Platen
  et~al.}{2007}]{Platen2007}
Platen E.,  {Van De Weygaert} R.,   Jones B. J.~T.,  2007, \mn@doi [Monthly
  Notices of the Royal Astronomical Society]
  {10.1111/j.1365-2966.2007.12125.x}, 380, 551

\bibitem[\protect\citeauthoryear{{Popescu}, {Tuffs}, {Dopita}, {Fischera},
  {Kylafis}  \& {Madore}}{{Popescu} et~al.}{2011}]{POPESCU2011}
{Popescu} C.~C.,  {Tuffs} R.~J.,  {Dopita} M.~A.,  {Fischera} J.,  {Kylafis}
  N.~D.,   {Madore} B.~F.,  2011, \mn@doi [\aap] {10.1051/0004-6361/201015217},
  \href {http://adsabs.harvard.edu/abs/2011A%26A...527A.109P} {527, A109}

\bibitem[\protect\citeauthoryear{Porter, Raychaudhury, Pimbblet  \&
  Drinkwater}{Porter et~al.}{2008}]{Porter2008}
Porter S.~C.,  Raychaudhury S.,  Pimbblet K.~A.,   Drinkwater M.~J.,  2008,
  Monthly Notices of the Royal Astronomical Society, 388, 1152

\bibitem[\protect\citeauthoryear{Ricciardelli, Cava, Varela  \&
  Quilis}{Ricciardelli et~al.}{2014}]{Ricciardelli2014}
Ricciardelli E.,  Cava A.,  Varela J.,   Quilis V.,  2014, \mn@doi [Monthly
  Notices of the Royal Astronomical Society] {10.1093/mnras/stu2061}, 445, 4045

\bibitem[\protect\citeauthoryear{Robotham \& Obreschkow}{Robotham \&
  Obreschkow}{2015}]{Robotham2015}
Robotham Â.,  Obreschkow Â.,  2015, eprint arXiv:1508.02145

\bibitem[\protect\citeauthoryear{Robotham et~al.,}{Robotham
  et~al.}{2011}]{Robotham2011}
Robotham A. S.~G.,  et~al., 2011, \mn@doi [Monthly Notices of the Royal
  Astronomical Society] {10.1111/j.1365-2966.2011.19217.x}, 416, 2640

\bibitem[\protect\citeauthoryear{Robotham et~al.,}{Robotham
  et~al.}{2013}]{Robotham2013a}
Robotham A. S.~G.,  et~al., 2013, \mn@doi [Monthly Notices of the Royal
  Astronomical Society] {10.1093/mnras/stt156}, 431, 167

\bibitem[\protect\citeauthoryear{Rojas, Vogeley, Hoyle  \& Brinkmann}{Rojas
  et~al.}{2004}]{Rojas2004}
Rojas R.~R.,  Vogeley M.~S.,  Hoyle F.,   Brinkmann J.,  2004, \mn@doi [The
  Astrophysical Journal] {10.1086/425225}, 617, 50

\bibitem[\protect\citeauthoryear{Rojas, Vogeley, Hoyle  \& Brinkmann}{Rojas
  et~al.}{2005}]{Rojas2005}
Rojas R.~R.,  Vogeley M.~S.,  Hoyle F.,   Brinkmann J.,  2005, \mn@doi [The
  Astrophysical Journal] {10.1086/428476}, 624, 571

\bibitem[\protect\citeauthoryear{{Salim} et~al.,}{{Salim}
  et~al.}{2007}]{SALIM2007}
{Salim} S.,  et~al., 2007, \mn@doi [\apjs] {10.1086/519218}, \href
  {http://adsabs.harvard.edu/abs/2007ApJS..173..267S} {173, 267}

\bibitem[\protect\citeauthoryear{{Salpeter}}{{Salpeter}}{1955}]{SALPETER1955}
{Salpeter} E.~E.,  1955, \mn@doi [\apj] {10.1086/145971}, \href
  {http://adsabs.harvard.edu/abs/1955ApJ...121..161S} {121, 161}

\bibitem[\protect\citeauthoryear{Schlegel, Finkbeiner  \& Davis}{Schlegel
  et~al.}{1998}]{Schlegel1998}
Schlegel D.~J.,  Finkbeiner D.~P.,   Davis M.,  1998, The Astrophysical
  Journal, 500, 525

\bibitem[\protect\citeauthoryear{Schwarz}{Schwarz}{1978}]{schwarz1978}
Schwarz G.,  1978, \mn@doi [Ann. Statist.] {10.1214/aos/1176344136}, 6, 461

\bibitem[\protect\citeauthoryear{Shandarin \& Zeldovich}{Shandarin \&
  Zeldovich}{1989}]{Shandarin1989}
Shandarin S.~F.,  Zeldovich Y.~B.,  1989, \mn@doi [Reviews of Modern Physics]
  {10.1103/RevModPhys.61.185}, 61, 185

\bibitem[\protect\citeauthoryear{{Snedden}, {Coughlin}, {Phillips}, {Mathews}
  \& {Suh}}{{Snedden} et~al.}{2016}]{Snedden2016}
{Snedden} A.,  {Coughlin} J.,  {Phillips} L.~A.,  {Mathews} G.,   {Suh} I.-S.,
  2016, \mn@doi [\mnras] {10.1093/mnras/stv2421}, \href
  {http://adsabs.harvard.edu/abs/2016MNRAS.455.2804S} {455, 2804}

\bibitem[\protect\citeauthoryear{Sousbie}{Sousbie}{2011}]{Sousbie2011}
Sousbie T.,  2011, \mn@doi [Monthly Notices of the Royal Astronomical Society]
  {10.1111/j.1365-2966.2011.18394.x}, 414, 350

\bibitem[\protect\citeauthoryear{Taylor et~al.,}{Taylor
  et~al.}{2011}]{Taylor2011}
Taylor E.~N.,  et~al., 2011, \mn@doi [Monthly Notices of the Royal Astronomical
  Society] {10.1111/j.1365-2966.2011.19536.x}, 418, 1587

\bibitem[\protect\citeauthoryear{Tempel, Stoica, Martinez, Liivamagi, Castellan
   \& Saar}{Tempel et~al.}{2014}]{Tempel2014}
Tempel E.,  Stoica R.~S.,  Martinez V.~J.,  Liivamagi L.~J.,  Castellan G.,
  Saar E.,  2014, \mn@doi [Monthly Notices of the Royal Astronomical Society]
  {10.1093/mnras/stt2454}, 438, 3465

\bibitem[\protect\citeauthoryear{Tuffs, Popescu, V�lk, Kylafis  \&
  Dopita}{Tuffs et~al.}{2004}]{Tuffs2004}
Tuffs R.~J.,  Popescu C.~C.,  V�lk H.~J.,  Kylafis N.~D.,   Dopita M.~A.,
  2004, \mn@doi [Astronomy and Astrophysics] {10.1051/0004-6361:20035689}, 419,
  821

\bibitem[\protect\citeauthoryear{Welker, Dubois, Devriendt, Pichon, Kaviraj  \&
  Peirani}{Welker et~al.}{2015}]{Welker2015}
Welker Â.,  Dubois Â.,  Devriendt Â.,  Pichon Â.,  Kaviraj Â.,   Peirani Â.,
  2015, eprint arXiv:1502.05053

\bibitem[\protect\citeauthoryear{{Whitaker}, {van Dokkum}, {Brammer}  \&
  {Franx}}{{Whitaker} et~al.}{2012}]{WHITAKER2012}
{Whitaker} K.~E.,  {van Dokkum} P.~G.,  {Brammer} G.,   {Franx} M.,  2012,
  \mn@doi [\apjl] {10.1088/2041-8205/754/2/L29}, \href
  {http://adsabs.harvard.edu/abs/2012ApJ...754L..29W} {754, L29}

\bibitem[\protect\citeauthoryear{Wijesinghe et~al.,}{Wijesinghe
  et~al.}{2012}]{Wijesinghe2012}
Wijesinghe D.~B.,  et~al., 2012, \mn@doi [Monthly Notices of the Royal
  Astronomical Society] {10.1111/j.1365-2966.2012.21164.x}, 423, 3679

\bibitem[\protect\citeauthoryear{{Wyder} et~al.,}{{Wyder}
  et~al.}{2007}]{WYDER2007}
{Wyder} T.~K.,  et~al., 2007, \mn@doi [\apjs] {10.1086/521402}, \href
  {http://adsabs.harvard.edu/abs/2007ApJS..173..293W} {173, 293}

\bibitem[\protect\citeauthoryear{Yang, Mo  \& van~den Bosch}{Yang
  et~al.}{2009}]{Yang2009}
Yang X.,  Mo H.~J.,   van~den Bosch F.~C.,  2009, \mn@doi [The Astrophysical
  Journal] {10.1088/0004-637X/695/2/900}, 695, 900

\bibitem[\protect\citeauthoryear{Zehavi et~al.,}{Zehavi
  et~al.}{2011}]{Zehavi2011}
Zehavi I.,  et~al., 2011, \mn@doi [The Astrophysical Journal]
  {10.1088/0004-637X/736/1/59}, 736, 59

\bibitem[\protect\citeauthoryear{Zel'dovich}{Zel'dovich}{1970}]{Zel'dovich1970}
Zel'dovich Â.,  1970, Astronomy and Astrophysics, 5

\bibitem[\protect\citeauthoryear{Zheng, Zehavi, Eisenstein, Weinberg  \&
  Jing}{Zheng et~al.}{2009}]{Zheng2009}
Zheng Z.,  Zehavi I.,  Eisenstein D.~J.,  Weinberg D.~H.,   Jing Y.~P.,  2009,
  \mn@doi [The Astrophysical Journal] {10.1088/0004-637X/707/1/554}, 707, 554

\bibitem[\protect\citeauthoryear{{Zinger}, {Dekel}, {Birnboim}, {Kravtsov}  \&
  {Nagai}}{{Zinger} et~al.}{2015}]{2015arXiv151005388Z}
{Zinger} E.,  {Dekel} A.,  {Birnboim} Y.,  {Kravtsov} A.,   {Nagai} D.,  2015,
  preprint, \href {http://esoads.eso.org/abs/2015arXiv151005388Z} {}
  (\mn@eprint {arXiv} {1510.05388})

\makeatother
\end{thebibliography}

\label{lastpage}

\end{document}